# Heat Transfer Coefficients of Moving Particle Beds from Flow-Dependent Particle Bed Thermal Conductivity and Near-Wall Resistance

Sarath R. Adapa[*1], Xintong Zhang[*1], Tianshi Feng[1], Jian Zeng[1], Ka Man Chung[2], Kevin J. Albrecht[3], Clifford K. Ho[3], Dimitri A. Madden[3], Renkun Chen[#1,2]

[1]Department of Mechanical and Aerospace Engineering, University of California San Diego, La Jolla, California 92093, United States

[2]Program in Materials Science and Engineering, University of California San Diego, La Jolla, California 92093, United States

[3]Concentrating Solar Technologies Department, Sandia National Laboratories, 1515 Eubank Blvd. SE, Albuquerque, New Mexico, 87123, United States

[*]These authors contributed equally to this work

[#]Corresponding Author: rkchen@ucsd.edu

**Abstract**

Accurate determination of heat transfer coefficients for flowing packed particle beds is essential to the design of particle heat exchangers, and other thermal and thermochemical processes. While such dense granular flows mostly fall into the well-known plug-flow regime, the discrete nature of granular materials alters the thermal transport processes in both the near-wall and bulk regions of flowing particle beds from their stationary counterparts. As a result, heat transfer correlations based on the stationary particle bed thermal conductivity could be inadequate for flowing particles in a heat exchanger. Most earlier works have achieved a reasonable agreement with experiments by treating granular heat transfer media as a plug-flow continuum with a near-wall thermal resistance in series. However, the thermophysical properties of the continuum were often obtained from measurements on stationary beds owing to the difficulty of flowing bed measurements. In this work, it was found that the properties of a stationary bed are highly sensitive to the method of particle packing and there is a decrease in the particle bed thermal conductivity and increase in the near-wall thermal resistance, measured as an effective air gap thickness, on the onset of particle flow. These variations in the thermophysical properties of stationary and flowing particle beds can lead to errors in heat transfer coefficient calculations. Therefore, the heat transfer coefficients for granular flows were calculated using experimentally determined flowing particle bed thermal conductivity and near-wall air gap for ceramic particles - CARBO CP 40/100 (mean diameter=275 μm), HSP 40/70 (404 μm) and HSP 16/30 (956 μm); at velocities of 5-15 mm·s$^{-1}$; and temperatures of 300-650 °C. The thermal conductivity and air gap values for CP 40/100 and HSP 40/70 were further used to calculate heat transfer coefficients across different particle bed temperatures and velocities for different parallel-plate heat exchanger dimensions. These calculations, which show good agreement with experiments in literature, can be used as a guide for heat exchanger designs.

**Introduction**

The heating and cooling of granular materials is a routine step in multiple industrial processes and their thermophysical properties have been extensively studied. For example, it is well understood that the thermal conductivity of bulk granular materials, i.e., packed particle beds, is dominated by gas conduction through the interstitial voids around particle contacts [1,2]. A good understanding of granular heat transfer is essential for high temperature applications, such as additive manufacturing, pebble-bed nuclear reactors, thermal energy storage, and thermochemical reactions. In some of these processes, heat transfer to the granular material happens indirectly through a heat exchanger (HEX). The recent use of solar-absorbing ceramic particles as the heat transfer fluid in concentrated solar power plants (CSP) is one such case [3]. As seen from the maturity and wide availability of granular material handling equipment being one of the advantages of particle based CSP, the "fluid mechanics" of granular fluids is well understood. However, heat transfer to high temperature flowing granular media is still an area of active research.

Particle bed heat exchangers which involve the gravity-driven flow of a dense particle bed are analogous to the simple hydrodynamic problem of a quasi-static dense granular flow between two vertical parallel plates [4]. In such flows the particle inertia is negligible, and particles interact through long-lasting frictional contacts. Depending on the wall roughness, there is a shear layer of ~10 particles which quickly transitions to a plug flow in the bulk [5]. In contrast, the particle bed in fluidized bed heat exchangers is more dilute and dominated by binary collisions. Thus, the heat transfer coefficients (HTCs) in packed particle beds are generally much lower than that of fluidized beds and margins for errors in HEX sizing are low [6]. It is well understood that in such confined particle beds, the wall disrupts particle packing and leads to a low packing density near the wall. As reported in the work of Yagi and Kunii [7], this low density region creates a thermal resistance to heat transfer from the wall to the particle bed. A later work by Sullivan and Sabersky [8] also observed this thermal resistance in flowing packed-particle beds. While granular media can be treated as fluids with a continuum approximation, the presence of this near-wall thermal resistance, which is absent in regular fluids, could significantly decrease heat transfer.

Some of the earliest experiments on heat transfer to granular flows were conducted by Brinn et al [9] on sand flowing down a constant wall-temperature tube. Based on their observation of a plug flow, they calculated the thermal conductivity of the flowing beds using a single-medium continuum approximation. In beds with similar packing density, they found that the flowing packed bed had a 10% lower thermal conductivity than the stationary packed bed. However, they had a reasonably good fit of their data using the continuum approximation despite not accounting for a near-wall resistance. In the study by Sullivan and Sabersky [8] they found that the particle size (a measure of the discreteness) had a considerable impact on the measured Nusselt number at velocities larger than 5 mm·s$^{-1}$. However, this discrete nature could be approximated by a continuum medium plus a particle-size-dependent near-wall thermal resistance. They modelled this resistance as an effective air film layer with a thickness of $0.085 \cdot d_\mathrm{p}$ between the wall and first layer of particles, where $d_\mathrm{p}$ is the particle diameter. While they acknowledged that the thickness would depend on the roughness of the wall, they neglected the lower near-wall density and assumed a uniform bed density. Since heat transfer across this air layer and the moving particle bed are in series, at low velocities, the continuum resistance dominates over this near-wall resistance. They speculated that the low velocities in the work of Brinn et al. [9] could be the reason the near-wall thermal resistance was not important. The presence of an effective gas gap layer with a thickness of $0.1 \cdot d_\mathrm{p}$ was also independently found by Denloye and Botterill [10]. They also noted that for beds with a similar mean particle diameter, the bed with the larger polydispersity had better heat transfer as it could pack more closely in the near-wall region. Natarajan and Hunt [11] extended the work on wall roughness, noting that in dense granular flows this air gap layer is a function of near-wall density and is larger for rougher walls. They fitted their data with Sullivan and Sabersky's [8] model with good agreement, finding a $0.09 \cdot d_\mathrm{p}$ and $0.114 \cdot d_\mathrm{p}$ thick air gap layer for smooth and rough walls (roughness on the order of $d_\mathrm{p}$), respectively. However, at high velocities >100mm·s$^{-1}$, they observed a decrease in heat transfer which they attributed to a further decrease in the near-wall density. Schlünder [12] used a first principles approach to quantify this near-wall region with a HTC for heat transfer from the wall to the particle bed surface . In a comparison with Sullivan and Sabersky's moving bed experiment results, he found his calculated near-wall resistance for a stationary bed to be 2-4 times lower. He correctly attributed this to particle bed expansion, which while not unique to Sullivan and Sabersky's test design, is a characteristic of dense granular flows known as dilatancy [13,14]. This leads to a reduction in the particle bed thermal conductivity and increase in the near-wall thermal resistance on the onset of particle flow.

Recently, there has been increased interest in the HTCs of vertical gravity-driven dense granular flows [15–18] and researchers have used various models [19–22] to understand their results. However, the thermal conductivity and packing density of a stationary packed bed are used and approximations of the near-wall region are not rigorous. Apart from the differences caused by flow, the stationary particle packing and subsequently their properties, are themselves highly sensitive to filling conditions [23,24]. This could lead to situations where the measured HTCs of particle HEXs were 30-75% lower than simulations with stationary bed properties [25]. Schlünder's work [12] precisely calculates heat transfer in the wall adjacent region but it requires knowledge of the wall contact area, which is difficult to measure for flowing or irregular polydisperse beds. Further, the physical presence of the wall permeates into the bed to the order of a few particle diameters [23,26] and also affects the bulk particle bed. While Natarajan and Hunt's [20] considered this in their kinetic theory analysis, they had to approximate the near-wall region due to a breakdown of the continuum approximation at the wall .

It is indeed very challenging to model granular flows, as the discrete nature of particles with point wall-contacts and the continuum assumptions involved in the plug-flow models are opposites. This difference is even more important as the channel size approaches a few particle diameters, and the bed becomes more discrete. Further, a lot of factors important to accurate modelling such as the bed density near the wall or particle contact networks in the bulk are difficult to measure in flowing beds. In an earlier work [13], the authors used Modulated Photothermal Radiometry (MPR), a frequency domain non-contact technique to extract bulk particle bed thermal conductivity ($k_{\text{eff}}$) and near-wall thermal resistance, modelled as an effective air gap layer ($D_{\text{air}}$) from measurements on flowing particle beds. The non-contact nature of the technique allowed flowing bed measurement as-is, without the introduction of flow disturbance from heaters and thermocouples. Measuring both $k_{\text{eff}}$ and $D_{\text{air}}$ in-situ, allows the calculation of HTCs with minimal assumptions or hard-to-obtain parameters.

In this work, it is shown that the "stationary bed" is not unique. There are significant variations in $k_{\text{eff}}$ and $D_{\text{air}}$ in different stationary beds, depending on the method of particle packing. It was also found that there is an increase in the $D_{\text{air}}$ and decrease in $k_{\text{eff}}$ on the onset of flow [13]. Both findings suggest the need to use the properties of flowing particle beds, rather than stationary ones, for calculating the HTCs of moving particle bed heat exchangers. Using the measured $k_{\text{eff}}$ and $D_{\text{air}}$ of flowing ceramic particles with mean diameters of 275 µm (CP 40/100), 404 µm (HSP 40/70) and 956 µm (HSP 16/30), HTCs were calculated in the temperature range of 300-650 °C and with different bed velocities. The average HTCs calculated ranged from 225-350 W·m$^{-2}$K$^{-1}$ for a parallel plate geometry with 5 mm spacing and 500 mm length. The measured $k_{\text{eff}}$ and $D_{\text{air}}$ of CP 40/100 and HSP 40/70 were fitted as a function of temperature and then used to predict the HTCs of their flowing beds for a range of velocities, temperatures, and parallel-plate channel depths. The calculated HTCs for different HEX parameters reported in literature agreed well with their corresponding experimental measurements. Compared to previous work on HTC modelling or simulations, these calculations are based on measured and fitted properties of flowing particles. The authors expect that the calculated HTCs and the methodology of obtaining them as described in this work could be a useful tool for researchers and engineers working in the area of granular flow HEXs and other heat transfer equipment.

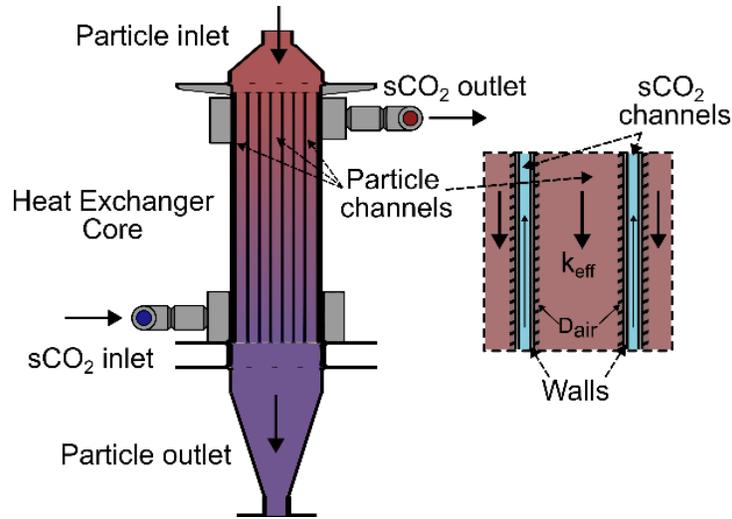

**Figure 1.** Schematic of a typical counterflowing parallel-plate packed particle bed heat exchanger. Close-up of the particle channel, heat exchange wall, and sCO₂ channel. The hatched region near the wall indicates the air gap layer.

**Experimental methods**

**Particle Media**

The granular media used in this work are polydisperse aluminosilicate ceramic particles manufactured by CARBO Ceramics, namely CARBOBEAD CP 40/100, CARBOBEAD HSP 40/70 and CARBOBEAD HSP 16/30 with mean diameters ($d_p$) of 275, 404, and 956 μm, respectively [27]. The presence of ferrite and other metal oxide additives provides an absorptivity of >0.9 suitable for CSP applications. Their mechanical and thermophysical properties in the stationary state have been well characterized in literature [2,28,29].

**Thermal Conductivity Measurements of Packed Particles**

The thermal conductivity of packed particles was measured in the stationary state using time-domain Transient Hot Wire (THW) [2], and both the stationary and flowing states with frequency-domain Modulated Photothermal Radiometry (MPR) [13]. The details of the measurement conditions and interpretation of the different packed particle bed thermal conductivities obtained are provided in this section.

*Flowing Particle Beds*

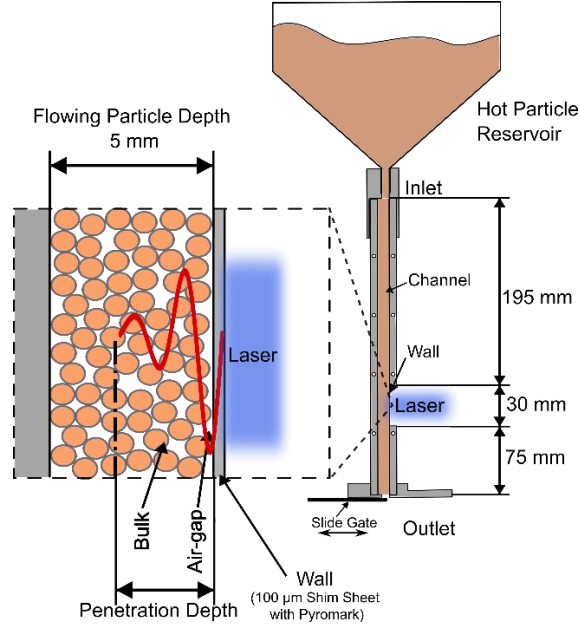

**Figure 2.** Schematic of flowing particle bed MPR experiments in a channel with 5 mm depth.

Fig. 2 shows a schematic of the flowing particle bed MPR experiments intended to replicate the conditions of a parallel-plate HEX with a 5 mm channel depth. A particle reservoir was used to heat and store particles at the desired experiment temperature. These particles flowed down a 300 mm × 30 mm × 5 mm channel under gravity and the flow velocity was controlled by a slide gate at the bottom outlet. The front wall of the particle channel (shown on the right side of the channel) was a 100 μm thick alloy shim sheet coated with a black light-absorbing paint. The frequency domain technique provides spatial resolution and allows the extraction of both the bulk thermal conductivity ($k_{\text{eff}}$) and near-wall air gap ($D_{\text{air}}$) information. More details about the MPR methodology and instrumentation are provided in Supplementary Information S1 and the reader is also referred to [13,30,31].

$k_{\text{eff}}$ and $D_{\text{air}}$ were measured for flowing particle beds of CP 40/100 ($d_{\text{p}}$= 275 μm) and HSP 40/70 ($d_{\text{p}}$= 404 μm) with velocities of 5, 10 and 15 mm·s$^{-1}$, typical of packed bed HEXs. The measurements were conducted at three nominal temperatures of 300, 500 and 650 °C. Due to some overlap in the particle size distribution of CP 40/100 and HSP 40/70, a third particle type, HSP 16/30 ($d_{\text{p}}$=956 μm) was also measured with a velocity of 12-15 mm·s$^{-1}$ in a similar temperature range. The flow of the large HSP 16/30 particles was not stable for velocities below 12 mm·s$^{-1}$, thus the data on velocities of 5 and 10 mm·s$^{-1}$ are not available. The measurements on HSP 16/30 also give insights into the behavior of flows with significantly larger particle size and a smaller channel depth to particle size ratio.

### Stationary Particle Beds

An important factor determining the packing structure and consequently the bed properties, is how the bed was made stationary or packed. Shown in the schematic in Fig. 3, are the four different stationary particle bed configurations measured. While the THW experiment measured $k_{\text{eff}}$ of dense particle beds, the MPR experiments measured both $D_{\text{air}}$ and $k_{\text{eff}}$ for different particle packings.

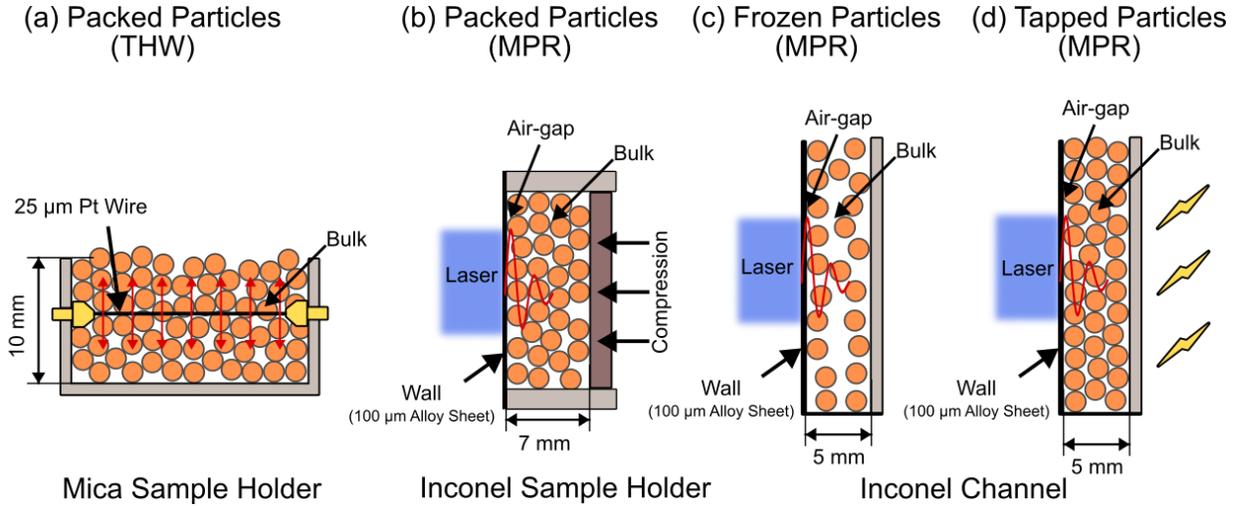

**Figure 3.** Schematic of different stationary particle bed thermal conductivity measurements. **(a)** Bulk transient hot wire. MPR measurements measuring both near-wall air gap layer and bulk particle bed regions - **(b)** Dense particle bed packed manually in a holder, **(c)** Frozen state and **(d)** Tapped state.

For the THW experiments (Fig. 3a), a particle bed was packed into a 10 mm deep mica container and made dense through vibration. A 25 μm-diameter platinum wire buried in the middle acts as both the heat source and thermometer. Since the wire is a line source conformal to the particle contours, this does not disturb packing or create any low-density regions. Similarly, the particles were also packed under vibration into an Inconel holder and held under compression for MPR experiments (Fig. 3b). The compression was needed to maintain the bed packing density in the vertical orientation and was slight enough to not create any deformation. The particle bed densities were roughly 2000 kg·m$^{-3}$, 2200 kg·m$^{-3}$ and 2350 kg·m$^{-3}$ for CP 40/100, HSP 40/70 and HSP 16/30, respectively, in both packed particle states (Figs. 3a & 3b). This corresponds to a packing density of 57%, 63% and 67%, respectively. The presence of the wall in the MPR experiments (Fig. 3b), similar to the flowing particle MPR setup, creates a disturbance in the particle packing and a near-wall $D_{air}$. Packed MPR and THW data for CP 40/100 and HSP 40/70 have previously been reported in [13] and [2].

After establishing a baseline $k_{eff}$ and $D_{air}$ on packed particles using THW and MPR, stationary bed MPR measurements were also conducted in the same channel used for flowing measurements (Figs. 3c & 3d). First, the slide gate at the outlet was completely opened, allowing new particles from the reservoir to enter the channel. Once the flow was stabilized, the gate was then suddenly closed, bringing the particles to a stop and letting them stack/pack from the bottom. This is called the 'frozen' state as this packing retains some characteristics of the flowing state [32] (Fig. 3c). Once this frozen state was measured, the channel was gently tapped with a hammer to densify the packing in the channel and this is referred to as the 'tapped' state (Fig. 3d) [33]. Granular systems share physical behavior with glasses and the kinetic energy of the individual particles, referred to as the granular temperature, is analogous to thermodynamic temperature. Thus, the making of the 'frozen' state is equivalent to the rapid quenching of a high temperature liquid into a glass. The perturbations induced by tapping are equivalent to thermal relaxation via annealing [34]. The density of the packed beds filled in the channel, was 1900 kg·m$^{-3}$, 2090 kg·m$^{-3}$ and 2300 kg·m$^{-3}$ for CP 40/100, HSP 40/70 and HSP 16/30, respectively. They correspond to packing densities of 54%, 60% and 66%, respectively. There was no significant difference in the densities between the tapped and frozen states for the same particle beds.

**Heat Transfer Coefficient (HTC) calculation**

The flowing particle bed is modelled as a plug-flow with a near-wall air gap layer, similar to the method of Sullivan and Sabersky [8], but also extend it to the fully developed region of flows confined between parallel plates. A schematic for the heat transfer model is shown in Fig. 4. The thermal resistance from the air gap is in series with the continuum thermal resistance of the moving bed and together, they give rise to the overall particle-to-wall HTC. In the model, the depth of the channel is $2b$ and the width is assumed to be infinite, justified by the much larger width compared to the depth and particle diameter [5] in HEXs. The boundary conditions in a real HEX are neither that of a constant wall temperature nor constant wall-flux but the constant wall flux condition can be used as the upper limit. Further, obtaining an analytical Nusselt number correlation for the constant wall-temperature condition is complicated as the interface between the air gap and plug flow of the particle bed cannot be defined with a simple boundary condition. Thus, in this work the constant wall heat flux condition was solved based on plug-flow analytical correlations in [35,36].

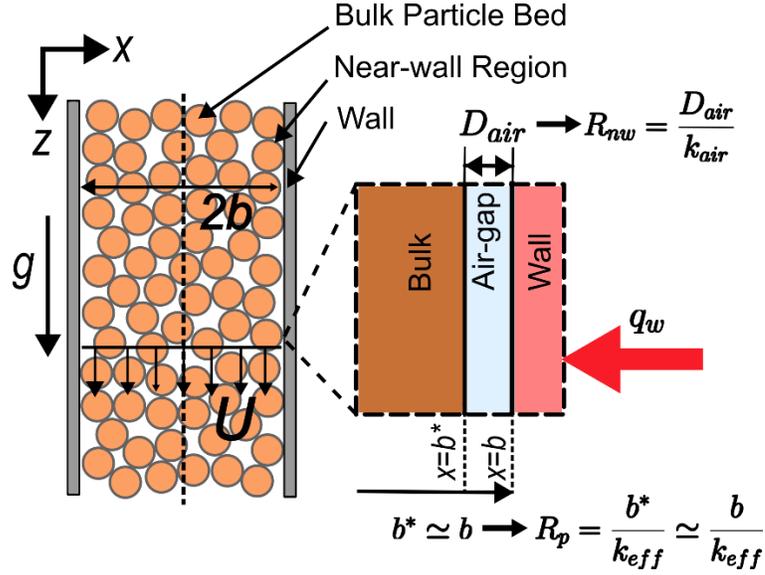

**Figure 4.** Schematic of the heat transfer model illustrating the bulk particle bed, near-wall region, and wall with constant heat flux.

For a parallel plate channel with a separation of $2b$, the hydraulic diameter is $D_h = 4b$. The near-wall resistance $R_{nw} = \frac{D_{air}}{k_{air}}$ is the thermal resistance of the effective air gap layer, where $k_{air}$ is the thermal conductivity of air at that temperature. Similarly, the resistance of the bulk particle bed is $R_p = \frac{b^*}{k_{eff}} \simeq \frac{b}{k_{eff}}$. The particle bed has thermal diffusivity $\alpha$ and it flows along the direction $z$ with a mean velocity $U$. The flowing particle bed is heated by a wall with constant flux $q_w$. The normalized difference between the temperature of the air gap/bulk particle bed interface and the mean bulk bed temperature, is then given by [35],

$$\frac{T(b^*,z) - T_m(z)}{\frac{q_w 4b}{k_{eff}}} = \frac{1}{12} - \frac{1}{2}\sum_{n=1}^{\infty} \frac{1}{n^2\pi^2} \exp\left(-\frac{n^2\pi^2 z\alpha}{b^2 U}\right) \tag{1}$$

The temperature drop across the air gap is given as,

$$T(b,z) - T(b^*,z) = q_w R_{nw} \tag{2}$$

Combining (1) and (2)

$$\frac{T(b,z) - T_\mathrm{m}(z)}{\frac{q_\mathrm{w} 4b}{k_\mathrm{eff}}} = \frac{1}{12} - \frac{1}{2}\sum_{n=1}^{\infty} \frac{1}{n^2\pi^2} \exp\left(-\frac{n^2\pi^2 z\alpha}{b^2 U}\right) + \frac{R_\mathrm{nw} k_\mathrm{eff}}{4b} \qquad (3)$$

The local Nusselt number and subsequently the local HTC can be obtained as,

$$Nu_{D_\mathrm{h}}(z) = \left[\frac{1}{12} - \frac{1}{2}\sum_{n=1}^{\infty} \frac{1}{n^2\pi^2} \exp\left(-\frac{16n^2\pi^2}{Gz}\right) + \frac{R_\mathrm{nw} k_\mathrm{eff}}{4b}\right]^{-1} \qquad (4)$$

$$h(z) = \frac{Nu_{D_\mathrm{h}} k_\mathrm{eff}}{D_\mathrm{h}} \qquad (5)$$

where $Gz$ is the dimensionless Graetz number given as

$$Gz = \frac{D_\mathrm{h}}{z} Pe_{D_\mathrm{h}} \qquad (6)$$

and $Pe_{D_\mathrm{h}}$ is the dimensionless Peclet number given as

$$Pe_{D_\mathrm{h}} = \frac{U D_\mathrm{h}}{\alpha} \qquad (7)$$

Averaging Eqn. 3 over a heated length $L$,

$$\overline{\frac{T(b,z) - T_\mathrm{m}(z)}{\frac{q_\mathrm{w} 4b}{k_\mathrm{eff}}}} = \frac{1}{L}\left[\frac{L}{12} + \frac{1}{2}\sum_{n=1}^{\infty} \frac{1}{n^4\pi^4} \frac{b^2 U}{\alpha}\left(\exp\left(-\frac{n^2\pi^2 L\alpha}{b^2 U}\right) - 1\right) + \frac{R_\mathrm{nw} k_\mathrm{eff} L}{4b}\right] \qquad (8)$$

And the average Nusselt number for a constant wall-flux plug-flow with a near-wall thermal resistance is given as,

$$\overline{Nu_{D_\mathrm{h}}} = \left[\frac{1}{12} + \sum_{n=1}^{\infty} \frac{1}{32n^4\pi^4} \frac{D_\mathrm{h}}{L} Pe_{D_\mathrm{h}}\left(\exp\left(-\frac{16n^2\pi^2 L}{D_\mathrm{h} Pe_{D_\mathrm{h}}}\right) - 1\right) + \frac{R_\mathrm{nw}}{4R_\mathrm{p}}\right]^{-1} \qquad (9)$$

From the Nusselt number, the average HTC can then be calculated as,

$$\bar{h} = \frac{\overline{Nu_{D_\mathrm{h}}} k_\mathrm{eff}}{D_\mathrm{h}} \qquad (10)$$

This method of using the measured $k_\mathrm{eff}$ and $D_\mathrm{air}$ of a flowing particle bed and applying the corresponding Nusselt number correlations to calculate HTCs is applicable to different flow configurations. Here, the discussion is limited to the simple case of flow in a parallel-plate channel and more details on other configurations can be found in [37].

**Results and Discussion**

**Stationary Particle Beds**

The thermal conductivity of different stationary particle beds was measured with both THW and MPR, as described earlier and shown in Fig. 3. In the configuration of the THW experiments, the thin platinum wire was placed in the center of the bulk region of the bed with particles pressed around it and thus, measuring the $k_{\text{eff}}$ of a fully packed bed. This was considered a dense bed with random packing and with absence of $D_{\text{air}}$ around the thin wire. In particle HEXs, where the heat transfer happens from a large surface to the particle bed, the presence of a tube or wall creates an adjacent air gap. The configurations of the MPR experiments better capture this behavior as the particle bed was confined behind a wall. Further, the heat diffuses from the wall, through the near-wall region and then into the bulk particle bed, closely resembling heat transfer in HEXs. The $k_{\text{eff}}$ from THW measurements, and the extracted $D_{\text{air}}$ and $k_{\text{eff}}$ from different MPR measurements for stationary particle beds are summarized in Table 1, 2 and 3, for CP 40/100, HSP 40/70 and HSP 16/30, respectively. Data at 650 °C is not available for the frozen and tapped configurations in the Inconel channel as it did not have sufficient heating power to reach the temperature.

Table 1: $k_{\text{eff}}$ and $D_{\text{air}}$ for stationary CP 40/100 particle beds.

| CP 40/100 $d_p$ =275 μm | 300 °C | | 500 °C | | 650 °C | |
| --- | --- | --- | --- | --- | --- | --- |
| | $k_{\text{eff}}$ (Wm$^{-1}$K$^{-1}$) | $D_{\text{air}}$ (μm) | $k_{\text{eff}}$ (Wm$^{-1}$K$^{-1}$) | $D_{\text{air}}$ (μm) | $k_{\text{eff}}$ (Wm$^{-1}$K$^{-1}$) | $D_{\text{air}}$ (μm) |
| Frozen | 0.21 | 18 | 0.35 | 16 | – | – |
| Tapped | 0.31 | 20 | 0.43 | 19 | – | – |
| Packed (MPR) | 0.29 | 14 | 0.34 | 14 | 0.37 | 16 |
| Packed (THW) | 0.39 | N/A | 0.43 | N/A | 0.47 | N/A |

Table 2: $k_{\text{eff}}$ and $D_{\text{air}}$ for stationary HSP 40/70 particle beds.

| HSP 40/70 $d_p$ =404 μm | 300 °C | | 500 °C | | 650 °C | |
| --- | --- | --- | --- | --- | --- | --- |
| | $k_{\text{eff}}$ (Wm$^{-1}$K$^{-1}$) | $D_{\text{air}}$ (μm) | $k_{\text{eff}}$ (Wm$^{-1}$K$^{-1}$) | $D_{\text{air}}$ (μm) | $k_{\text{eff}}$ (Wm$^{-1}$K$^{-1}$) | $D_{\text{air}}$ (μm) |
| Frozen | 0.26 | 30 | 0.27 | 27 | – | – |
| Tapped | 0.33 | 26 | 0.35 | 24 | – | – |
| Packed (MPR) | 0.30 | 19 | 0.36 | 21 | 0.39 | 22 |
| Packed (THW) | 0.38 | N/A | 0.46 | N/A | 0.50 | N/A |

Table 3: $k_{\text{eff}}$ and $D_{air}$ for stationary HSP 16/30 particle beds.

| HSP 16/30 $d_p$ =956 μm | 350 °C | | 500 °C | | 650 °C | |
| --- | --- | --- | --- | --- | --- | --- |
| | $k_{\text{eff}}$ (Wm$^{-1}$K$^{-1}$) | $D_{\text{air}}$ (μm) | $k_{\text{eff}}$ (Wm$^{-1}$K$^{-1}$) | $D_{\text{air}}$ (μm) | $k_{\text{eff}}$ (Wm$^{-1}$K$^{-1}$) | $D_{\text{air}}$ (μm) |
| Frozen | 0.53 | 99 | 0.75 | 107 | – | – |
| Tapped | 0.64 | 110 | 0.66 | 89 | – | – |
| Packed (MPR) | 0.53 | 86 | 0.79 | 99 | 0.68 | 98 |
| Packed (THW*) | 0.55 | N/A | 0.71 | N/A | 0.75 | N/A |

*The THW data for HSP 16/30 is from a 100% $N_2$ ambient instead of air (78% $N_2$).

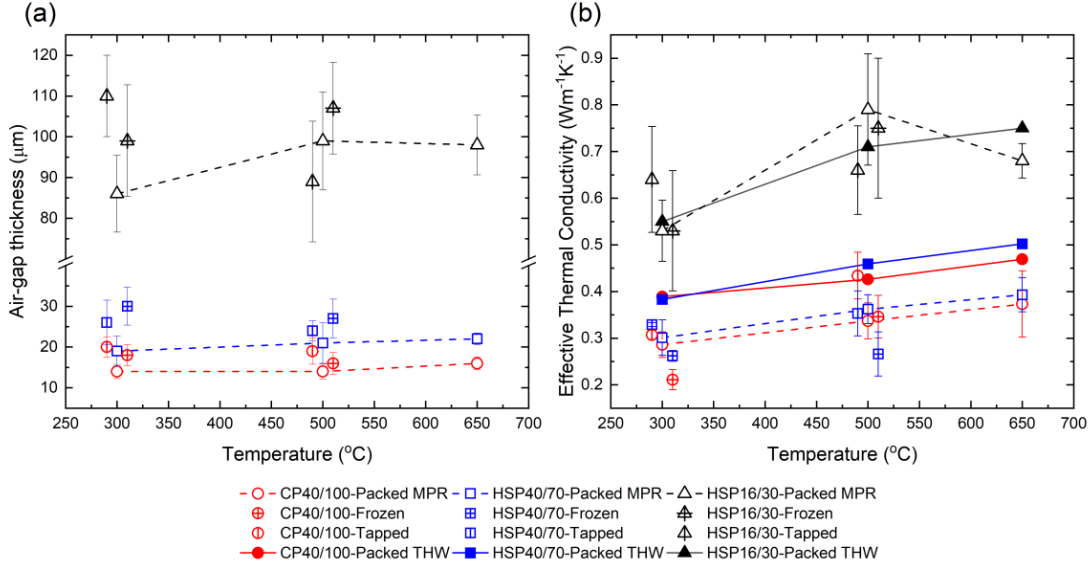

**Figure 5. (a)** Near-wall air gap thickness ($D_{air}$) from different packing methods for CP 40/100, HSP 40/70 and HSP 16/30. There is no air gap in the THW measurements. **(b)** Effective particle bed thermal conductivity ($k_{eff}$) from different packing methods for CP 40/100, HSP 40/70 and HSP 16/30. Error bars for $k_{eff}$ and $D_{air}$ is uncertainty calculated by the methodology described in Supplementary Information S2. The data points at 300 °C and 500 °C have been offset about their nominal temperatures to improve readability. Packed MPR and THW data for CP 40/100 and HSP 40/70 have previously been reported in [13] and [2].

As seen from Fig. 5, for the smaller particles, while all methods of packing indicate an increase in $k_{eff}$ with temperature, there is a large amount of scatter between methods for both $k_{eff}$ and $D_{air}$. The standard deviation of $k_{eff}$ is $13 - 19\%$ of mean and $10 - 15\%$ of mean for CP 40/100 and HSP 40/70, respectively. Similarly, the standard deviation for $D_{air}$ is also quite large, with $16 - 17\%$ of mean and $13 - 22\%$ of mean for CP 40/100 and HSP 40/70, respectively. For HSP 16/30, the scatter is at $< 9\%$ for $k_{eff}$ and $< 12\%$ for $D_{air}$, due to the higher baseline $k_{eff}$ and $D_{air}$ from the larger particle diameter. As illustrated in Fig. 3, both MPR and THW measurements would span $\sim 5 - 7 \cdot d_p$ for HSP 16/30. While this is treated as a bulk packed particle bed, the increased discreteness might skew some trends. Further, due to experimental challenges with high temperature measurements, the data for packed particle MPR for HSP 16/30 was from the fitting of a single measurement while all other measurements were conducted in triplicates, with fresh packings for each 300 to 650 °C measurements. Thus, the error bar for packed MPR on HSP16/30 is from fitting error only. Similarly, the THW experiments for HSP16/30 had to be done in an inert $N_2$ atmosphere to prevent rapid oxidation of the electrical connections.

As mentioned earlier, the packing structure and subsequently, the thermal conductivity of packed beds is highly sensitive to the method of filling [32,33]. It can be noted that the $k_{eff}$ from all packing methods in the MPR measurements are generally lower than THW measurements. While the frequency modulation of MPR allows a change in the probed length by changing the thermal penetration depth, the maximum depth (~1 mm) at the lowest frequency (0.03 Hz) is still within 3 particles from the wall, which could be affected by the near-wall region [13]. Further, in narrow channels, where the depth is on the order of a few particle diameters, the wall could affect the entire particle bed. Hypothetically, the low $k_{eff}$ from the MPR measurements could be caused by lower particle packing density of the particles in the near-wall region accessible by MPR. This is also supported by the observation that tapping the channel leads to compaction of the bed [33], leading to decreased $D_{air}$ and increased $k_{eff}$ as revealed in the MPR experiments. Importantly, this finding clearly demonstrates the large uncertainties and variation in the thermal conductivity of stationary beds, which is critically dependent on the method of packing and thus cannot be used for calculating the HTC of flowing particle beds. In particular, stationary particle bed thermal conductivity measurement techniques such as the THW method, which has the highest value and hence the largest deviation from those of flowing particle beds, would cause the largest error for HTC calculation. Therefore, it becomes apparent that using thermophysical properties of flowing particle beds is required for accurately calculating HTCs of particle bed HEXs.

**Flowing Particle Beds**

Complete details on the flowing particle bed MPR measurements on CP 40/100 (275 μm) and HSP40/70 (404 μm) have been reported in our earlier paper [13]. Since the average size of these particles was similar due to some overlap in the size distribution, a larger particle, HSP 16/30 (956 μm) was additionally measured in the flowing state. The summary of the $k_{eff}$ and $D_{air}$ results are given in Tables 4, 5 and 6.

**Table 4**: $k_{eff}$ and $D_{air}$ for flowing CP 40/100 particle beds. This data has previously been reported in [13].

| CP 40/100 $d_p$ =275 μm | 350 °C | | 460 °C | | 650 °C | |
| --- | --- | --- | --- | --- | --- | --- |
| | $k_{eff}$ (Wm$^{-1}$K$^{-1}$) | $D_{air}$ (μm) | $k_{eff}$ (Wm$^{-1}$K$^{-1}$) | $D_{air}$ (μm) | $k_{eff}$ (Wm$^{-1}$K$^{-1}$) | $D_{air}$ (μm) |
| 5 mm s$^{-1}$ | 0.21 | 29 | 0.29 | 31 | 0.30 | 34 |
| 10 mm s$^{-1}$ | 0.22 | 31 | 0.28 | 27 | 0.31 | 32 |
| 15 mm s$^{-1}$ | 0.22 | 31 | 0.29 | 30 | 0.31 | 33 |

**Table 5**: $k_{eff}$ and $D_{air}$ for flowing HSP 40/70 particle beds. This data has previously been reported in [13].

| HSP 40/70 $d_p$ =404 μm | 300 °C | | 480 °C | | 650 °C | |
| --- | --- | --- | --- | --- | --- | --- |
| | $k_{eff}$ (Wm$^{-1}$K$^{-1}$) | $D_{air}$ (μm) | $k_{eff}$ (Wm$^{-1}$K$^{-1}$) | $D_{air}$ (μm) | $k_{eff}$ (Wm$^{-1}$K$^{-1}$) | $D_{air}$ (μm) |
| 5 mm s$^{-1}$ | 0.27 | 28 | 0.31 | 32 | 0.32 | 33 |
| 10 mm s$^{-1}$ | 0.26 | 25 | 0.28 | 35 | 0.31 | 31 |

| 15 mm s$^{-1}$ | 0.26 | 29 | 0.27 | 36 | 0.29 | 30 |

Table 6: $k_{\text{eff}}$ and $D_{\text{air}}$ for flowing HSP 16/30 particle beds.

| HSP 16/30 $d_p$ =956 µm | 325 °C | | 450 °C | | 600 °C | |
| --- | --- | --- | --- | --- | --- | --- |
| | $k_{\text{eff}}$ (Wm$^{-1}$K$^{-1}$) | $D_{\text{air}}$ (µm) | $k_{\text{eff}}$ (Wm$^{-1}$K$^{-1}$) | $D_{\text{air}}$ (µm) | $k_{\text{eff}}$ (Wm$^{-1}$K$^{-1}$) | $D_{\text{air}}$ (µm) |
| 12-15 mm s$^{-1}$ | 0.41 | 88 | 0.57 | 99 | 0.59 | 118 |

As seen from Fig. 6, the flowing air gap thickness measured is roughly $(0.08 - 0.12) \cdot d_p$ of the particle diameter for all three particle sizes and agrees with the mean literature values of $\sim 0.1 \cdot d_p$ [8,10,11]. This holds valid despite the small channel depth to particle size ratio of HSP 16/30 (*5.23*) compared to CP 40/100 (*18.18*) and HSP 40/70 (*12.5*). However, with decreasing channel sizes, the granular flow can no longer be considered a packed particle bed. Further, there are concerns of clogging when the channel depth is less than about $10 \cdot d_p$ [38]. While particle beds with larger particles have a significantly higher thermal conductivity, they also have high near-wall air gaps and require the use of wider channels (parallel-plate spacing). This leads to increased channel depth and associated HEX costs, and we limit our later discussion to the smaller particles CP 40/100 and HSP 40/70.

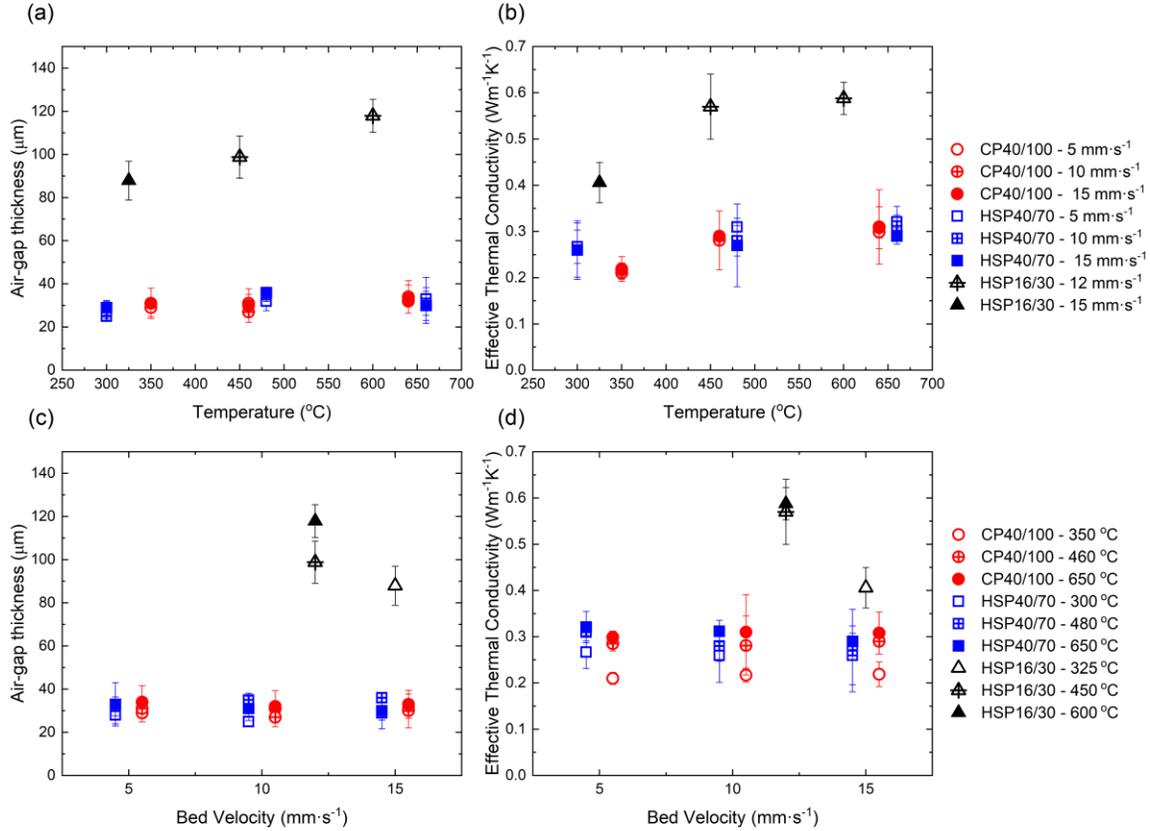

**Figure 6.** Near-wall air gap thickness ($D_{air}$) and effective particle bed thermal conductivity ($k_{eff}$) of flowing particle beds as a function of **(a, b)** temperature and **(c, d)** velocity. Error bars for $k_{eff}$ and $D_{air}$ is uncertainty calculated by the methodology described in Supplementary Information S2. In panels **(a,b)** overlapping data points for CP 40/100 and HSP 40/70 at 650 °C have been offset about their nominal temperature and in panels **(c,d)** data points for CP 40/100 and HSP 40/70 at 5, 10 and 15 mm·s$^{-1}$ have been offset about their nominal velocities to improve readability. This data has previously been reported in [13].

It can be seen from Fig. 6 (d), that the error in $k_{eff}$ increases with increasing velocity, which is likely from increasing uncertainty in the flow rate itself, but overall, it is relatively constant. However, still within the limits of error, it is expected that $k_{eff}$ is sensitive and will increase with temperature. This can be understood by the physics behind $k_{eff}$, where gas conduction has a major contribution, and $k_{air}$ itself increases with temperature [2]. $D_{air}$ on the other hand, is modelled as a physical air layer and should be independent of temperature. However, in Fig. 6 (a) $D_{air}$ of HSP 16/30 increases with temperature. It is possible that due to the large air gap between the wall and particle bed from the larger particle size, there is a larger radiation component. The model used to fit MPR data in this work [13] (and Supplementary Information S1) considers diffusive heat transfer from the wall to the particle bed and radiation from the wall which is also present in the frequency range for the air gap, could be missed. Similarly, the effective air gap in Sullivan and Sabersky's theory [8] also neglects radiation. A complete understanding of radiative heat transfer in large diameter particle beds is still required thus the parameter fitting is limited to the smaller CP 40/100 and HSP 40/70 particles. There is no clear trend observed in either $k_{eff}$ or $D_{air}$ with increasing the particle bed flow velocity and both are mostly constant. It is likely that within the velocity range of 5-15 mm·s$^{-1}$, which is typical for particle bed HEXs, the flow is still dense and the velocities are not large enough to see a decrease in overall heat transfer, as observed in other works [11]. However, apart from the experiment and fitting error, the polydispersity of the individual particle media and the overlap in particle sizes between them could also obfuscate some finer trends.

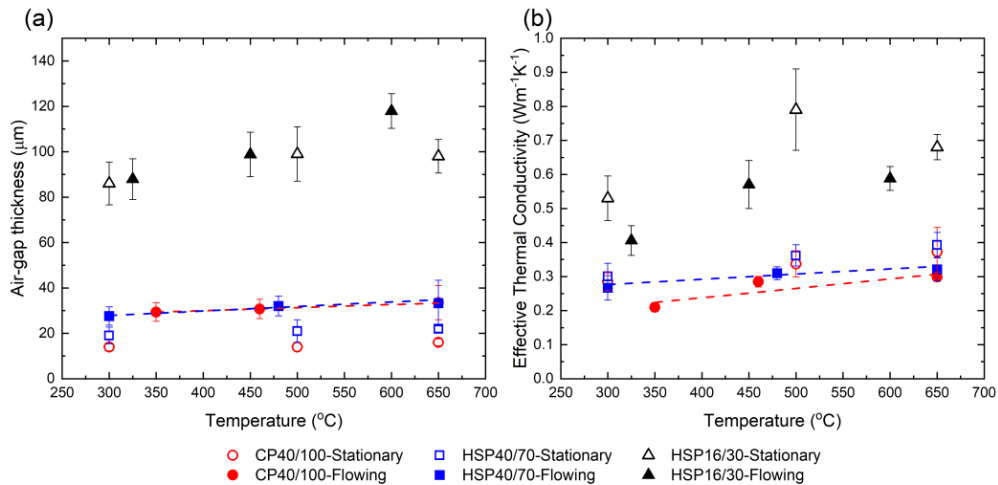

**Figure 7. (a)** Near-wall air gap thickness ($D_{air}$) and **(b)** Effective particle bed thermal conductivity ($k_{eff}$), for stationary and flowing CP 40/100, HSP 40/70, and HSP 16/30 as a function of temperature. The stationary data is from MPR measurements on packed particle beds discussed earlier (Fig. 3(b)). The flowing points are properties at the lowest stable velocities (5 mm·s$^{-1}$ for CP40/100, HSP40/70 and 12-15 mm·s$^{-1}$ for HSP16/30). Error bars for $k_{eff}$ and $D_{air}$ is uncertainty calculated by the methodology described in Supplementary Information S2. The dashed lines are linear fits for flowing CP 40/100 and HSP 40/70

properties as a function of temperature, see Table 7. The data for CP 40/70 and HSP 40/70 has been previously reported in [13].

Subsequently, $k_{eff}$ and $D_{air}$ are considered at 5 mm·s$^{-1}$ for CP40/100 and HSP40/70 and 12-15 mm·s$^{-1}$ for HSP16/30, due to the lower error from lower flow rate uncertainty. This is compared with the properties of MPR on the packed particle bed configuration discussed earlier and shown in Fig. 3(b). It is apparent that there is a decrease in the $k_{eff}$ and an increase in the $D_{air}$ of the flowing particle beds. On the onset of flow, $k_{eff}$ decreases by 15-27% for CP 40/100 and 11-18% HSP 40/70, respectively. The air gap thickness also increases from roughly $0.05 \cdot d_p$ in the stationary bed to $(0.08 - 0.12) \cdot d_p$ for both particles. However, for HSP16/30, while $D_{air}$ remained fairly constant between stationary and flowing beds, the $k_{eff}$ also decreased by 14-26%.

It was found that the onset of flow is accompanied by dilatancy, reducing the number of particle contacts which consequently increases the size of the interstitial void around the particle contact and decreases the bed thermal conductivity [13]. The flowing particles also have higher velocity fluctuations near the wall which widens the near-wall air gap thickness [13,20,39,40]. However, the flow and their collisions with the wall continuously introduces new particle bed packing structures as it enters the channel, leading to a more average but uniform flowing $k_{eff}$ and $D_{air}$ at a particular temperature. The flowing bed parameters are fit as linear functions of temperature (dashed lines in Fig. 7) and later used to interpolate both $k_{eff}$ and $D_{air}$ for different HEX parameter analysis. Details of the fitting lines are given below in Table 7.

**Table 7**: Linear fits of $k_{eff}$ and $D_{air}$ for flowing particle beds as $y = m \times T + c$ where T is temperature in °C. Data and fitting lines are shown in Fig. 7.

| Particle | Interpolated Property | Slope - m | Intercept - c |
|---|---|---|---|
| CP 40/100 | $D_{air}$ (μm) | $1.3 \times 10^{-2}$ | 25 |
| $d_p$ =275 μm | $k_{eff}$ (W·m$^{-1}$K$^{-1}$) | $2.8 \times 10^{-4}$ | 0.13 |
| HSP 40/70 | $D_{air}$ (μm) | $2 \times 10^{-2}$ | 22 |
| $d_p$ =404 μm | $k_{eff}$ (W·m$^{-1}$K$^{-1}$) | $1.5 \times 10^{-4}$ | 0.23 |

**Heat Transfer Calculations from measured $k_{eff}$ and $D_{air}$**

To highlight the errors that can occur from using stationary particle bed properties, the HTCs are calculated for different ($k_{eff}$, $D_{air}$) cases. The cases are as follows: (i) Stationary packed particle bed $k_{eff}$ from THW and neglecting near-wall thermal resistance or $D_{air} = 0$. This is typical of measurements where the particle bed and sensor geometries are different from the flowing particle channels. (ii) Stationary bed $k_{eff}$ and $D_{air}$ from MPR measurements. While most works understand the presence of a near-wall thermal resistance and include it in their analysis, they often base the values of both $k_{eff}$ and $D_{air}$ from stationary particle beds. As shown so far, flowing beds have different properties. (iii) Flowing bed $k_{eff}$ and $D_{air}$ from MPR measurements. This is the case where the properties used for heat transfer calculations are directly measured in the flowing state. Local HTCs were calculated for constant wall-flux heat transfer to a plug-flow continuum and incorporating a near-wall thermal resistance, as described in the methodology earlier. Results for the three cases are plotted from Eqns. 4-5 as a function of the inverse Graetz number ($Gz^{-1} = \frac{z}{D_h Pe_{D_h}}$) in Fig. 8.

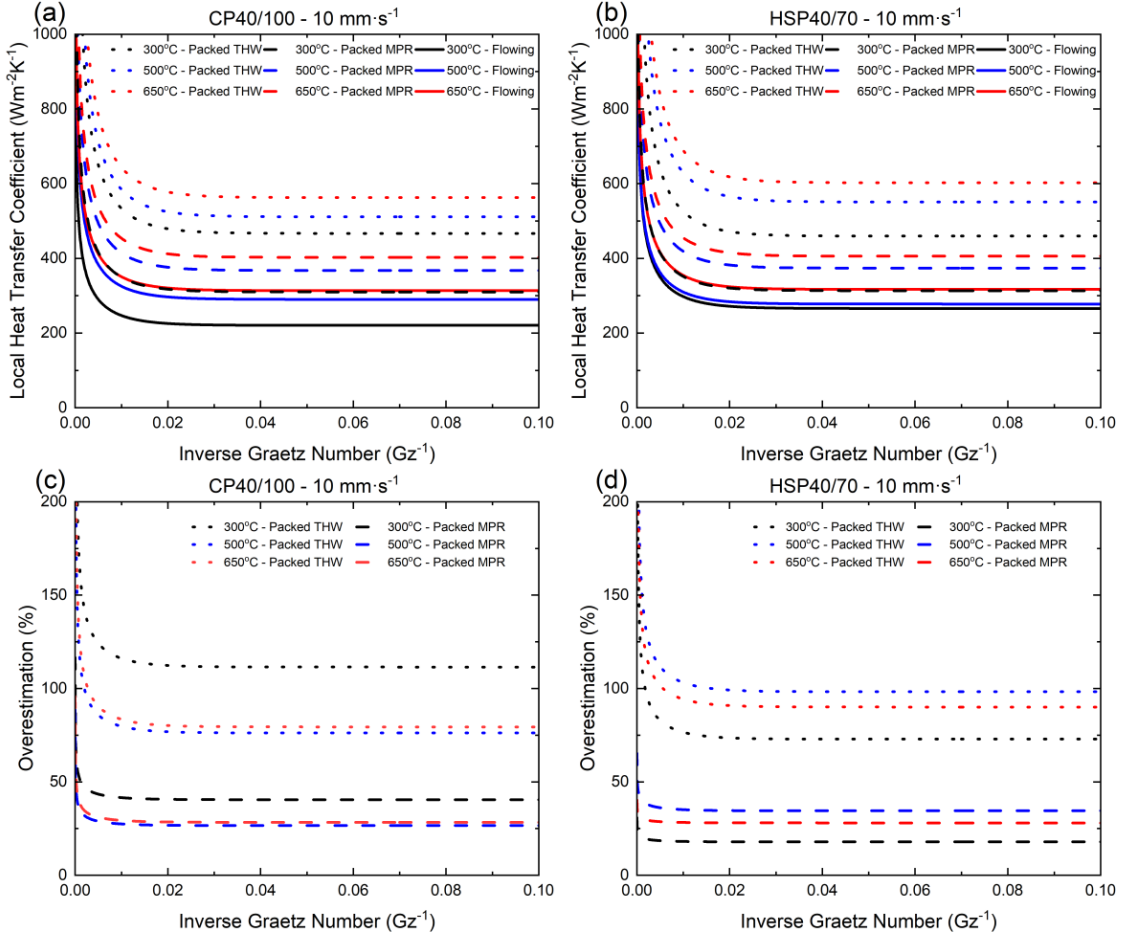

**Figure 8.** Local HTCs as a function of inverse Graetz number for a particle bed velocity of 10 mm·s$^{-1}$ obtained from Eqns. 4-5 **(a)** CP 40/100, **(b)** HSP 40/70. Solid lines are calculated using $k_{eff}$ and $D_{air}$ obtained from flowing MPR measurements. Dashed lines are calculated using $k_{eff}$ and $D_{air}$ obtained from packed particle bed MPR measurements in the Inconel holder. Dotted lines are calculated using $k_{eff}$ from packed particle THW data and no air gap. **(c)** and **(d)** show the corresponding percentage of overestimation over the flowing particle bed HTCs.

While the entrance region has large continuum HTCs, the presence of the near-wall resistance can significantly reduce the combined HTC. For example, the HTC obtained for flowing beds at 650 °C is about the same if the properties for 300 °C stationary beds were used for calculations. It would indeed come as a surprise if the expected improvement in HTC by the increase in $k_{eff}$ going from 300 to 650 °C, would vanish when the particle bed starts to flow. As seen from Fig. 8 (c) and (d), if $k_{eff}$ measured from THW was used to calculate HTC without accounting for any $D_{air}$ near the wall, the HTC is overestimated in the fully developed region by $75 - 110\%$. And even if this was corrected by attempting to measure in a wall confined configuration with corresponding $k_{eff}$ and $D_{air}$, the overestimation is still around $25 - 40\%$.

A typical parallel-plate HEX has a length of 500 mm and channel depths (plate spacing) of a few millimeters. The thermal entrance length for dense granular flows in such HEXs is on the order of a few centimeters and the flow is thermally fully developed for most of the channel length. Fig. 9 (a) shows the calculated average Nusselt numbers as a function of Peclet number for a channel with 5 mm depth and 500 mm length and measured $k_{eff}$ and $D_{air}$. $\overline{Nu_{D_h}}$ increases slightly with increasing $Pe_{D_h}$ as the entrance region also increases. For both CP 40/100 and HSP 40/70, this lies between 10-12. However, for HSP 16/30 which has much larger air gap thickness, $\overline{Nu_{D_h}}$ is around 6. Eqn. 4 and 9 were modified to the fully developed case below, where $Nu_{D_h}$ is only a function of the ratios of near-wall thermal resistance from the air gap layer to bulk particle bed thermal resistance from the particle bed in the channel.

$$\overline{Nu_{D_h}} = \left(\frac{1}{12} + \frac{R_{nw}}{4R_p}\right)^{-1} \qquad (11)$$

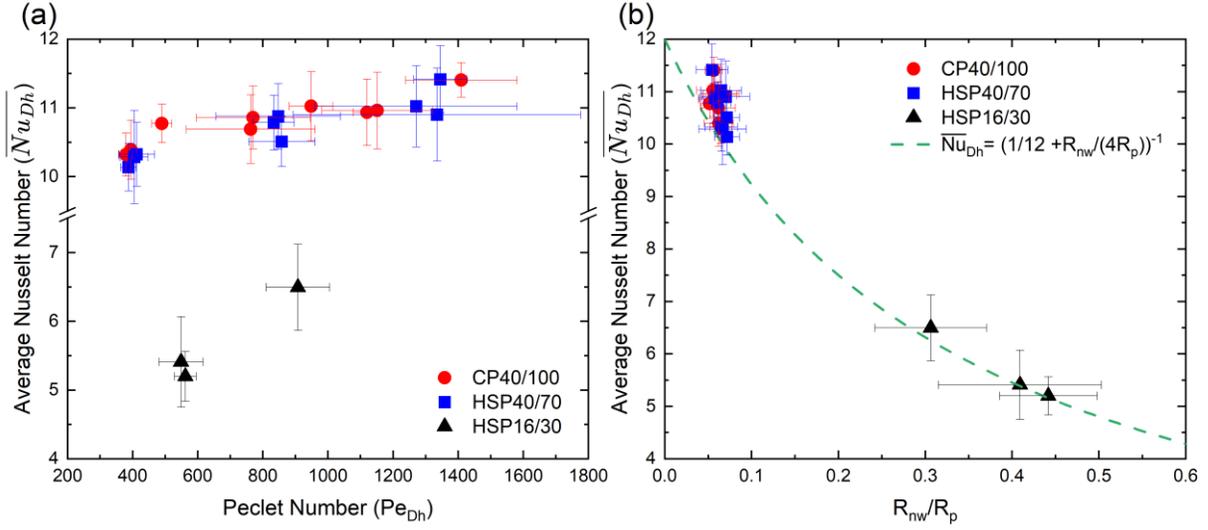

**Figure 9.** Calculations for a parallel-plate channel with a 5 mm depth and 500 mm length. **(a)** Average Nusselt number as a function of Peclet number; **(b)** Average Nusselt number as a function of the ratios of near-wall thermal resistance from air gap to bulk particle bed thermal resistance. Dashed line is Eqn. 11. Error bars are calculated based on the error propagation of $k_{eff}$ and $D_{air}$.

The average HTCs calculated from Eqn. 10, are shown in Fig. 10. The values lie between 225-350 W·m$^{-2}$K$^{-1}$ with an uncertainty ≤25% and agree well with the typical values for high-temperature flowing particle bed HTCs [15]. Despite the low $\overline{Nu_{D_h}}$ from the large air gap in the HSP 16/30 particle beds, the high particle bed thermal conductivity leads to similar HTCs as the other particles. However, larger particles would only be advantageous over the smaller particles in applications requiring large channel depths.

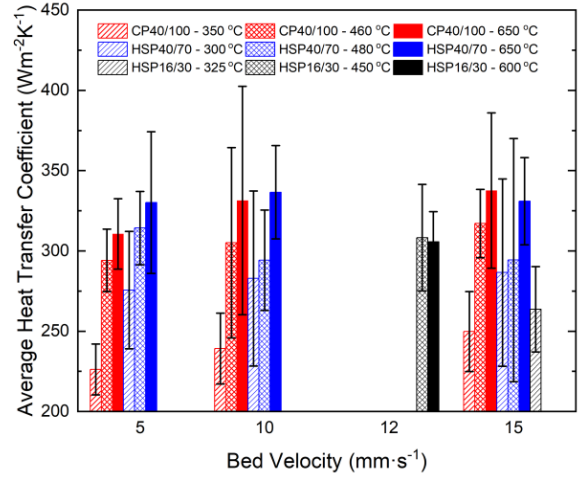

**Figure 10.** HTCs calculated for a parallel-plate channel with a 5 mm depth and 500 mm length. Error bars are calculated based on the error propagation of $k_{\text{eff}}$ and $D_{\text{air}}$.

### Heat Transfer Calculations from fitted $k_{\text{eff}}$ and $D_{\text{air}}$

As described earlier, in long heat exchanger channels, for a fixed particle size there is little variation in Nusselt number due to the fully developed nature of flow. While both terms in the $R_{\text{nw}}/R_{\text{p}}$ ratio depend on $k_{\text{air}}$ which is a function of particle bed temperature, as seen in Fig. 9, the effect on Nusselt number is negligible. Thus, from Eqn. 10, HTC is inversely proportional to the channel depth ($D_{\text{h}}/2$) and directly proportional to the effective particle bed conductivity ($k_{\text{eff}}$). Typically, the channel depth (plate spacing) needs to be larger than $10 \cdot d_{\text{p}}$ to prevent issues with particle clogging. The operating temperature which determines $k_{\text{eff}}$ is also limited by the material considerations and often limited to ~650 °C in the current CSP applications [32] and ~700 °C for the Gen3 particle pilot plant [3]. In Fig. 11, Eqns. 9-10 are used to calculate HTCs over a temperature range of 300-650 °C and channel depths of 3-10 mm for particle bed velocities of 5-15 mm·s⁻¹. Both $k_{\text{eff}}$ and $D_{\text{air}}$ were obtained from the linear fitting functions of temperature given in Table 7 and assumed to be independent of velocity in the dense granular flow regime.

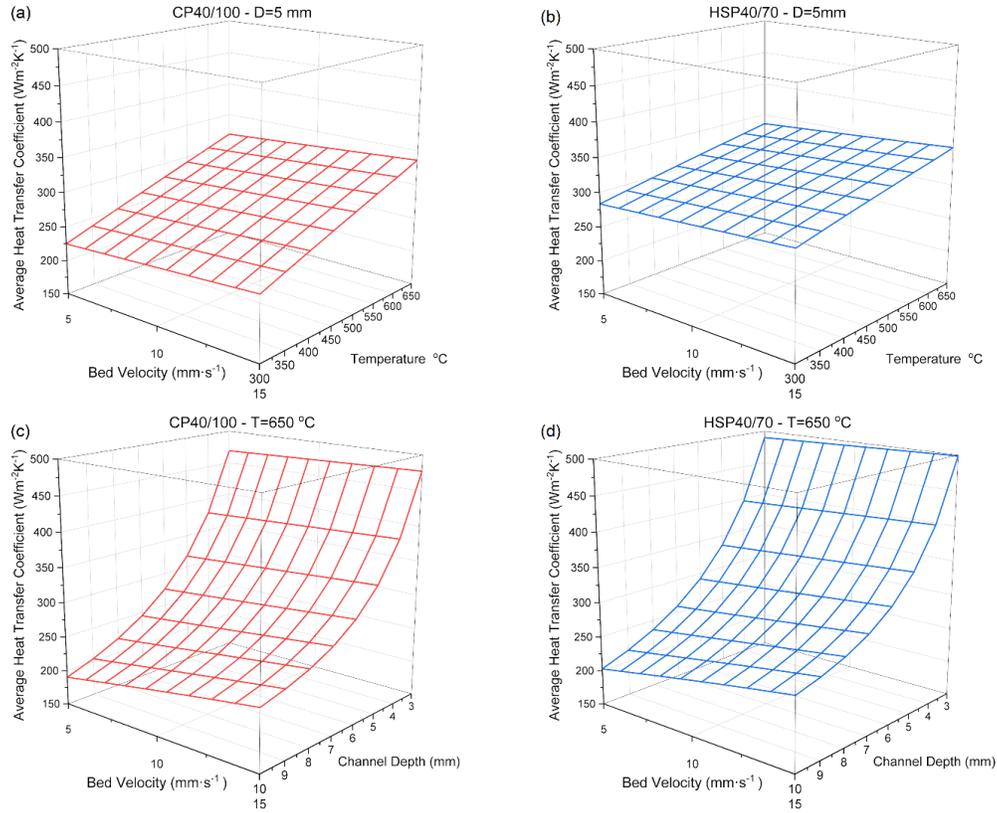

**Figure 11.** Average HTCs for CP 40/100 and HSP 40/70 in a 500 mm long channel for different bed velocities and **(a, b)** temperature; **(c, d)** channel depth.

At a temperature of 650 °C, a 3 mm channel depth can lead to HTCs close to 500 W·m$^{-2}$K$^{-1}$ and practical for most high-temperature industrial processes. A comparison is made to the work of Albrecht et al. [41] who measured the overall HTC (U$_{HEX}$) of a 20 kW$_{th}$ sCO$_2$-particle bed HEX using HSP 40/70 as the particle media. Their HEX had a channel depth of 3 mm and length of 465 mm. The particle mass flow rate reported in their work was converted to velocity based on the channel dimensions and the packed bed density. They varied the particle velocities from 13-38 mm·s$^{-1}$ and the sCO$_2$ flow rate from 50 g·s$^{-1}$ to 150 g·s$^{-1}$. As seen from Fig. 12, the sCO$_2$ flow rate has a large impact on HTC. It must be noted that their data is the overall HTC (U$_{HEX}$) which includes the thermal resistances of particle-to-wall, wall itself, and the wall-to-sCO$_2$ all in series. However, due to unknowns associated with the sCO$_2$ channel geometry the wall-to-sCO$_2$ HTC is not separated out. Their data shows that while HTC increases with particle temperature, it is insensitive to the particle flow velocity and supports the finding of near constant $k_{eff}$ and $D_{air}$ at different flow velocities in the dense particle bed regime. The particle-to-wall HTC is calculated using Eqns. 9-10 based on fitted $k_{eff}$ and $D_{air}$, and is shown as the solid line in Fig. 12. Since the calculated HTC only presents the particle-to-wall portion, it should represent the upper limit of all the measured overall HTC (U$_{HEX}$), which indeed is the case as shown in Fig. 12, thus validating the current work's methodology.

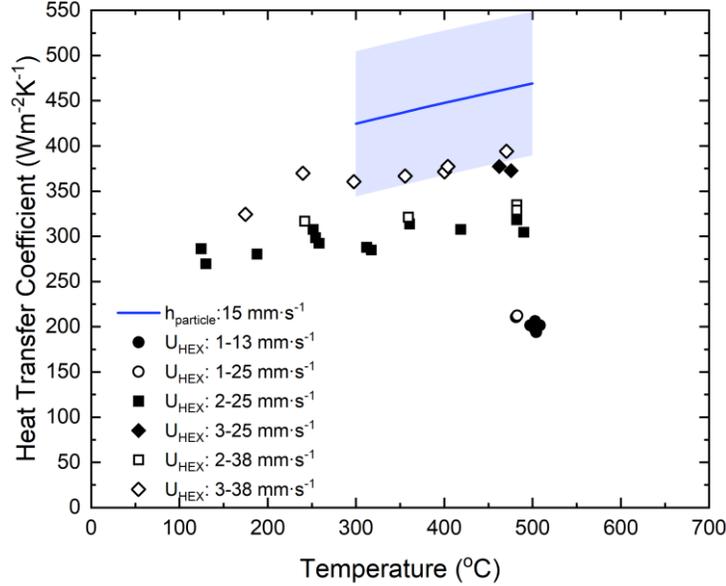

**Figure 12.** Comparison of calculated particle-to-wall HTCs with measured overall HTCs from [41]. The particle media was HSP 40/70, and the channel was 3 mm deep and 500 mm long. 1: 50 g·s$^{-1}$, 2: 100 g·s$^{-1}$ and 3: 150 g·s$^{-1}$ are different flow rates of sCO$_2$. The particle mass flow rates have been converted to velocities. The solid line is the calculated particle-to-wall HTC using Eqns. 9-10 with fitted $k_{eff}$ and $D_{air}$ from Table 7. Shaded region is error calculated based on the error propagation of $k_{eff}$ and $D_{air}$ of HSP40/70.

**Discussion**

    High HTC values will lead to a lower cost for moving particle bed heat exchangers which meet the Department of Energy's Gen3 LCOE goals [6]. It is known that a smaller channel depth can lead to a higher HTC for a continuum flow. However, for granular media, a minimum depth of around $10 \cdot d_p$ is recommended, below which the flow can be unstable or even clogged [38]. For HSP 40/70 and CP 40/100, this depth is around 3-5 mm and HTCs around 500 W·m$^{-2}$K$^{-1}$ at 650 °C were obtained in this work (Fig. 11). Another parameter that can be tuned is the size of the particle media. It is also well known that particle beds with larger particles have higher $k_{eff}$, especially from the larger radiation contribution at high temperatures. For HSP40/70, radiation was ~10% of overall conductivity at >600 °C [2]. However, as shown here, there exist conflicting effects of particle size on $k_{eff}$ and $D_{air}$: even though $k_{eff}$ is higher for particle beds with larger particles, the $D_{air}$ is also increased, as in the case of HSP 16/30 vs. HSP 40/70 and CP 40/100. The resultant overall effect on HTC with large particle size could be detrimental when the channel depth is small. Given this tradeoff, it is highly desirable if the low $D_{air}$ of small particles and the high $k_{eff}$ of large particles could be combined into one particle bed. Recent work by Stout et al. [24] for stationary particle beds showed that a bimodal particle distribution achieved through mixing HSP 40/70 and HSP 16/30 could achieve a larger thermal conductivity than either of the individual particles. They reported an enhancement of 20% and 30% at 21 °C and 300 °C, respectively. It would be interesting to examine if these beneficial effects can be retained at high temperature with flowing particles. Other approaches such as the use of internal finned structures [42], have already shown improvements of over 35% in HTC over non-finned tubes and mechanical vibration [43] has shown HTCs of ~1100 W·m$^{-2}$K$^{-1}$, comparable to the 1000 W·m$^{-2}$K$^{-1}$ readily achieved in fluidized beds [44]. However, the detailed mechanisms of heat transfer in fluidized/vibrated beds,

such as the thermal resistances associated with the near-wall air gap and the convection in the bulk bed, have yet to be fully understood and there could still be room for optimization. On this note, more innovative research into improving heat transfer through high thermal conductivity media or other techniques to reduce the near-wall air gap is still required to develop high performance particle HEXs.

**Conclusion**

In this work, the bulk thermal conductivity ($k_{\text{eff}}$) and near-wall resistance, in the form of an effective air gap ($D_{\text{air}}$), were reported for high temperature stationary and flowing particle beds. The bulk thermal conductivity and air gap of stationary particle beds are sensitive to the packing method and can have significant variation. The largest scatter in $k_{\text{eff}}$ of stationary beds was around 19% and 15% for HSP40/70 (mean diameter=404 μm) and CP40/100 (275 μm), respectively. Similarly, the scatter in $D_{\text{air}}$ of stationary beds was around 17% and 22% for HSP40/70 and CP40/100, respectively. While the scatter in HSP16/30 (956 μm) values was much lower, it had both a large $k_{\text{eff}}$ and $D_{\text{air}}$ due to the larger particle size. Further, on the onset of flow, there is a decrease in thermal conductivity (~11-27%) and an increase in air gap from 5% of particle diameter to around 8-12%. Thus, it is essential to determine $k_{\text{eff}}$ and $D_{\text{air}}$ based on in-situ measurements of flowing particle beds.

HTCs were calculated based on the measured $k_{\text{eff}}$ and $D_{\text{air}}$ for flowing beds of CP 40/100, HSP 40/70 and HSP 16/30 at velocities of 5-15 mm·s$^{-1}$ and temperatures of 300-650 °C. The HTCs of these flowing packed beds are in the range of 225-350 W·m$^{-2}$K$^{-1}$ with uncertainty ≤ 25% and agree well with literature values. Due to the suitability of 275 and 404 μm particles for HEX applications, their $k_{\text{eff}}$ and $D_{\text{air}}$ were fit as a linear function of temperature to calculate HTCs over the entire temperature range of 300-650 °C for different channel depths. This is justified by the physical dependence of $k_{\text{eff}}$ on gas thermal conductivity and the calculated HTCS were in good agreement with HEX experiments reported in literature. The HTC values and calculation methodology reported in the work can be used as a reference for particle heat exchanger design.

**Acknowledgement**

This paper is based upon work supported by the U.S. Department of Energy's Office of Energy Efficiency and Renewable Energy (EERE) under Solar Energy Technologies Office (SETO) Agreement Numbers DE-EE0008379 and DE-EE0009825. The views expressed herein do not necessarily represent the views of the U.S. Department of Energy or the United States Government.

**Nomenclature**

CSP – Concentrated Solar Power Plant

HEX – Heat Exchanger

HTC – Heat Transfer Coefficient

MPR – Modulated Photothermal Radiometry

THW – Transient Hot Wire

sCO$_2$ – Super-critical Carbon dioxide

**Symbols**

$b$ – Half of the channel width (m)

$b^*$ – Distance of air gap layer from center $= b - D_\text{air}$ (m)

$c_\text{p}$ – Specific Heat (J · kg$^{-1}$K$^{-1}$)

$D_\text{air}$ – Effective air gap thickness (μm)

$D_\text{h}$ – Hydraulic diameter of the channel $= 4b$ (m)

$d_\text{p}$ – Average particle diameter (μm)

$g$ – gravity acceleration (9.81 m · s$^{-1}$)

$Gz$ – Graetz number $= Pe_{D_\text{h}} D_\text{h}/z$

$h$ - Heat transfer coefficient (W · m$^{-2}$K$^{-1}$)

$\bar{h}$ - Average heat transfer coefficient (W · m$^{-2}$K$^{-1}$)

$k$ – Thermal conductivity (W · m$^{-1}$K$^{-1}$)

$k_\text{air}$ – Thermal conductivity of air (W · m$^{-1}$K$^{-1}$)

$k_\text{eff}$ – Effective bulk particle bed thermal conductivity (W · m$^{-1}$K$^{-1}$)

$L$ – Length of channel (m)

$L_\text{p}$ – Thermal penetration depth (m)

$Nu_{D_\text{h}}$ – Nusselt number $= hD_\text{h}/k_\text{eff}$

$\overline{Nu_{D_\text{h}}}$ – Averaged Nusselt number $= \bar{h}D_\text{h}/k_\text{eff}$

$Pe$ – Peclet number $= UD_\text{h}/\alpha$

$q_0$ – Laser Heat Flux (W · m$^{-2}$)

$q_\text{w}$ – Constant wall-heat flux (W · m$^{-2}$)

$R_\text{nw}$ – Near-wall thermal resistance (K · m$^2$W$^{-1}$) $= D_\text{air}/k_\text{air}$

$R_\text{p}$ – Bulk particle thermal resistance (K · m$^2$W$^{-1}$) $= b/k_\text{eff}$

$t$ – Time (s)

$T$ – Temperature (K)

$T_\text{m}$ – Mean bulk temperature of flowing particle bed (K)

$U$ – Particle bed flow velocity (m · s$^{-1}$)

$U_\text{HEX}$ – Overall heat exchanger heat transfer coefficient (W · m$^{-2}$K$^{-1}$)

$z$ – Direction of particle flow velocity (m)

**Greek Letters**

$\alpha$ – Thermal Diffusivity ($\text{m}^2\text{s}^{-1}$)

$\rho$ – Density ($\text{kg} \cdot \text{m}^{-3}$)

**References**

[1] van Antwerpen, W., du Toit, C. G., and Rousseau, P. G., 2010, "A Review of Correlations to Model the Packing Structure and Effective Thermal Conductivity in Packed Beds of Mono-Sized Spherical Particles," Nucl. Eng. Des., **240**(7), pp. 1803–1818.
[2] Chung, K. M., Zeng, J., Adapa, S. R., Feng, T., Bagepalli, M. V., Loutzenhiser, P. G., Albrecht, K. J., Ho, C. K., and Chen, R., 2021, "Measurement and Analysis of Thermal Conductivity of Ceramic Particle Beds for Solar Thermal Energy Storage," Sol. Energy Mater. Sol. Cells, **230**, p. 111271.
[3] "Generation 3 Concentrating Solar Power Systems (Gen3 CSP)," Energy.gov [Online]. Available: https://www.energy.gov/eere/solar/generation-3-concentrating-solar-power-systems-gen3-csp. [Accessed: 10-May-2024].
[4] GDR MiDi, 2004, "On Dense Granular Flows," Eur. Phys. J. E, **14**(4), pp. 341–365.
[5] Nedderman, R. M., and Laohakul, C., 1980, "The Thickness of the Shear Zone of Flowing Granular Materials," Powder Technol., **25**(1), pp. 91–100.
[6] Ho, C. K., Carlson, M., Albrecht, K. J., Ma, Z., Jeter, S., and Nguyen, C. M., 2019, "Evaluation of Alternative Designs for a High Temperature Particle-to-sCO2 Heat Exchanger," J. Sol. Energy Eng., **141**(2), p. 021001.
[7] Yagi, S., and Kunii, D., 1960, "Studies on Heat Transfer near Wall Surface in Packed Beds," AIChE J., **6**(1), pp. 97–104.
[8] Sullivan, W. N., and Sabersky, R. H., 1975, "Heat Transfer to Flowing Granular Media," Int. J. Heat Mass Transf., **18**(1), pp. 97–107.
[9] Brinn, M. S., Friedmen, S. J., Gluckert, F. A., and Pigford, R. L., 1948, "Heat Transfer to Granular Matgerials," Ind. Eng. Chem., **40**(6), pp. 1050–1061.
[10] Denloye, A. O. O., and Botterill, J. S. M., 1977, "Heat Transfer in Flowing Packed Beds," Chem. Eng. Sci., **32**(5), pp. 461–465.
[11] Natarajan, V. V. R., and Hunt, M. L., 1997, "Heat Transfer in Vertical Granular Flows," Exp. Heat Transf., **10**(2), pp. 89–107.
[12] Schlünder, E. U., 1984, "Heat Transfer to Packed and Stirred Beds from the Surface of Immersed Bodies," Chem. Eng. Process. Process Intensif., **18**(1), pp. 31–53.
[13] Zhang, X., Adapa, S., Feng, T., Zeng, J., Chung, K. M., Ho, C., Albrecht, K., and Chen, R., 2024, "Micromechanical Origin of Heat Transfer to Granular Flow," Phys. Rev. E, **109**(4), p. L042902.
[14] Reynolds, O., 1885, "On the Dilatancy of Media Composed of Rigid Particles in Contact. With Experimental Illustrations," Lond. Edinb. Dublin Philos. Mag. J. Sci., **20**(127), pp. 469–481.
[15] Watkins, M. F., and Gould, R. D., 2019, "Experimental Characterization of Heat Transfer to Vertical Dense Granular Flows Across Wide Temperature Range," J. Heat Transf., **141**(3), p. 032001.
[16] Yu, Y., Nie, F., Bai, F., and Wang, Z., 2021, "Theoretical and Experimental Investigation on Heating Moving Packed Beds in a Single Tube with Constant Wall Temperature," Int. J. Heat Mass Transf., **180**, p. 121725.
[17] Wei, G., Huang, P., Pan, L., Cui, L., Xu, C., and Du, X., 2022, "Experimental, Numerical and Analytical Modeling of Heat Transfer of Gravity Driven Dense Particle Flow in Vertical Heated Plates," Int. J. Heat Mass Transf., **187**, p. 122571.
[18] Maskalunas, J., Nellis, G., and Anderson, M., 2022, "The Heat Transfer Coefficient Associated with a Moving Packed Bed of Silica Particles Flowing through Parallel Plates," Sol. Energy, **234**, pp. 294–303.
[19] Botterill, J. S. M., and Denloye, A. O. O., 1978, "A Theoretical Model of Heat Transfer to a Packed or Quiescent Fluidized Bed," Chem. Eng. Sci., **33**(4), pp. 509–515.
[20] Natarajan, V. V. R., and Hunt, M. L., 1998, "Kinetic Theory Analysis of Heat Transfer in Granular Flows," Int. J. Heat Mass Transf., **41**(13), pp. 1929–1944.

[21] Albrecht, K. J., and Ho, C. K., 2019, "Heat Transfer Models of Moving Packed-Bed Particle-to-sCO2 Heat Exchangers," J. Sol. Energy Eng., **141**(3), p. 031006.
[22] Watkins, M. F., Chilamkurti, Y. N., and Gould, R. D., 2019, "Analytic Modeling of Heat Transfer to Vertical Dense Granular Flows," J. Heat Transf., **142**(2).
[23] Landry, J. W., Grest, G. S., Silbert, L. E., and Plimpton, S. J., 2003, "Confined Granular Packings: Structure, Stress, and Forces," Phys. Rev. E, **67**(4), p. 041303.
[24] Stout, D., Kandadai, N., and Otanicar, T. P., 2023, "Experimental and Numerical Investigation of Bimodal Particle Distributions for Enhanced Thermal Conductivity in Particle Based Concentrating Solar Power Applications," Sol. Energy, **263**, p. 111992.
[25] Carlson, M., Albrecht, K., Ho, C., Laubscher, H., and Alvarez, F., 2020, *High-Temperature Particle Heat Exchanger for sCO2 Power Cycles [Award 30342]*, SAND2020-14357, SuNLaMP-0000000–1507, 1817287.
[26] Desmond, K. W., and Weeks, E. R., 2009, "Random Close Packing of Disks and Spheres in Confined Geometries," Phys. Rev. E, **80**(5), p. 051305.
[27] "CARBOBEAD High-Performance Ceramic Media | CARBO" [Online]. Available: https://carboceramics.com/products/filtration/ceramic-filter-media/carbobead-product-detail. [Accessed: 10-Mar-2023].
[28] Siegel, N. P., Gross, M. D., and Coury, R., 2015, "The Development of Direct Absorption and Storage Media for Falling Particle Solar Central Receivers," J. Sol. Energy Eng., **137**(4).
[29] Bagepalli, M. V., Yarrington, J. D., Schrader, A. J., Zhang, Z. M., Ranjan, D., and Loutzenhiser, P. G., 2020, "Measurement of Flow Properties Coupled to Experimental and Numerical Analyses of Dense, Granular Flows for Solar Thermal Energy Storage," Sol. Energy, **207**, pp. 77–90.
[30] Zeng, J., Chung, K. M., Wang, Q., Wang, X., Pei, Y., Li, P., and Chen, R., 2021, "Measurement of High-Temperature Thermophysical Properties of Bulk and Coatings Using Modulated Photothermal Radiometry," Int. J. Heat Mass Transf., **170**, p. 120989.
[31] Zeng, J., Chung, K. M., Adapa, S. R., Feng, T., and Chen, R., 2021, "In-Situ Thermal Transport Measurement of Flowing Fluid Using Modulated Photothermal Radiometry," Int. J. Heat Mass Transf., **180**, p. 121767.
[32] Bertho, Y., Giorgiutti-Dauphiné, F., and Hulin, J.-P., 2003, "Dynamical Janssen Effect on Granular Packing with Moving Walls," Phys. Rev. Lett., **90**(14), p. 144301.
[33] Hong, D. C., Yue, S., Rudra, J. K., Choi, M. Y., and Kim, Y. W., 1994, "Granular Relaxation under Tapping and the Traffic Problem," Phys. Rev. E, **50**(5), pp. 4123–4135.
[34] Gago, P. A., and Boettcher, S., 2020, "Universal Features of Annealing and Aging in Compaction of Granular Piles," Proc. Natl. Acad. Sci., **117**(52), pp. 33072–33076.
[35] Muzychka, Y. S., Walsh, E., and Walsh, P., 2010, "Simple Models for Laminar Thermally Developing Slug Flow in Noncircular Ducts and Channels," J. Heat Transf., **132**(11).
[36] Kakaç, S., Shah, R. K., and Aung, W., eds., 1987, *Handbook of Single-Phase Convective Heat Transfer*, Wiley, New York.
[37] Zeng, J., Chung, K. M., Zhang, X., Feng, T., Adapa, S., and Chen, R., 2022, "Characterization and Understanding of Thermal Transport in Stationary and Moving Particle Beds for Concentrating Solar Power," Annu. Rev. Heat Transf., **25**(1), pp. 117–173.
[38] Albrecht, K. J., and Ho, C. K., 2019, "Design and Operating Considerations for a Shell-and-Plate, Moving Packed-Bed, Particle-to-sCO2 Heat Exchanger," Sol. Energy, **178**, pp. 331–340.
[39] Natarajan, V. V. R., Hunt, M. L., and Taylor, E. D., 1995, "Local Measurements of Velocity Fluctuations and Diffusion Coefficients for a Granular Material Flow," J. Fluid Mech., **304**, pp. 1–25.
[40] Jeong, S. Y., Bagepalli, M. V., Brooks, J. D., Ranjan, D., Zhang, Z. M., and Loutzenhiser, P. G., 2023, "Experimental and Numerical Analyses of Gravity-Driven Granular Flows between Vertical Parallel Plates for Solar Thermal Energy Storage and Transport," Int. J. Heat Mass Transf., **216**, p. 124571.
[41] Albrecht, K., Laubscher, H., Bowen, C., and Ho, C., 2022, *Performance Evaluation of a Prototype Moving Packed-Bed Particle/sCO2 Heat Exchanger.*, SAND2022-12615, 1887943, 709939.
[42] Jurtz, N., Flaischlen, S., Scherf, S. C., Kraume, M., and Wehinger, G. D., 2020, "Enhancing the Thermal Performance of Slender Packed Beds through Internal Heat Fins," Processes, **8**(12), p. 1528.

[43] Siebert, A. W., Highgate, D., and Newborough, M., 1999, "Heat Transfer Characteristics of Mechanically-Stimulated Particle Beds," Appl. Therm. Eng., **19**(1), pp. 37–49.

[44] Fosheim, J. R., Hernandez, X., Abraham, J., Thompson, A., Jesteadt, B., and Jackson, G. S., 2022, "Narrow-Channel Fluidized Beds for Particle-sCO2 Heat Exchangers in next Generation CPS Plants," AIP Conf. Proc., **2445**(1), p. 160007.

# Supplementary Information

# Heat Transfer Coefficients of Moving Particle Beds from Flow-Dependent Particle Bed Thermal Conductivity and Near-Wall Resistance

Sarath R. Adapa[*,1], Xintong Zhang[*,1], Tianshi Feng[1], Jian Zeng[1], Ka Man Chung[2], Kevin J. Albrecht[3], Clifford K. Ho[3], Dimitri A. Madden[3], Renkun Chen[#,1,2]

[1]Department of Mechanical and Aerospace Engineering, University of California San Diego, La Jolla, California 92093, United States

[2]Program in Materials Science and Engineering, University of California San Diego, La Jolla, California 92093, United States

[3]Concentrating Solar Technologies Department, Sandia National Laboratories, 1515 Eubank Blvd. SE, Albuquerque, New Mexico, 87123, United States

[*]These authors contributed equally to this work

[#]Corresponding Author: rkchen@ucsd.edu

**Contents**

## S1. MPR Method and Data Fitting

A summary of the Modulated Photothermal Radiometry (MPR) method, as described in [1] and used to obtain the effective particle bed thermal conductivity ($k_{\text{eff}}$) and near-wall thermal resistance as an air-gap ($D_{\text{air}}$) for this work, is provided below.

MPR is a frequency-domain non-contact measurement technique in which a modulated surface-heat-source establishes an oscillating temperature field inside the bulk sample. Decreasing the modulation frequency of the heat source increases the thermal penetration depth (and vice versa), which is given by,

$$L_p \propto \sqrt{\frac{\alpha}{\omega}} \tag{S1}$$

where $\alpha = \frac{k}{\rho C}$ is the thermal diffusivity of the sample ($k$ is the thermal conductivity, $\rho$ is the density and $C$ is the specific heat) and $\omega$ is the frequency in rad/s. The surface temperature changes as a function of the frequency, in response to the subsurface thermal effusivities ($e = \sqrt{k\rho C}$) [2].

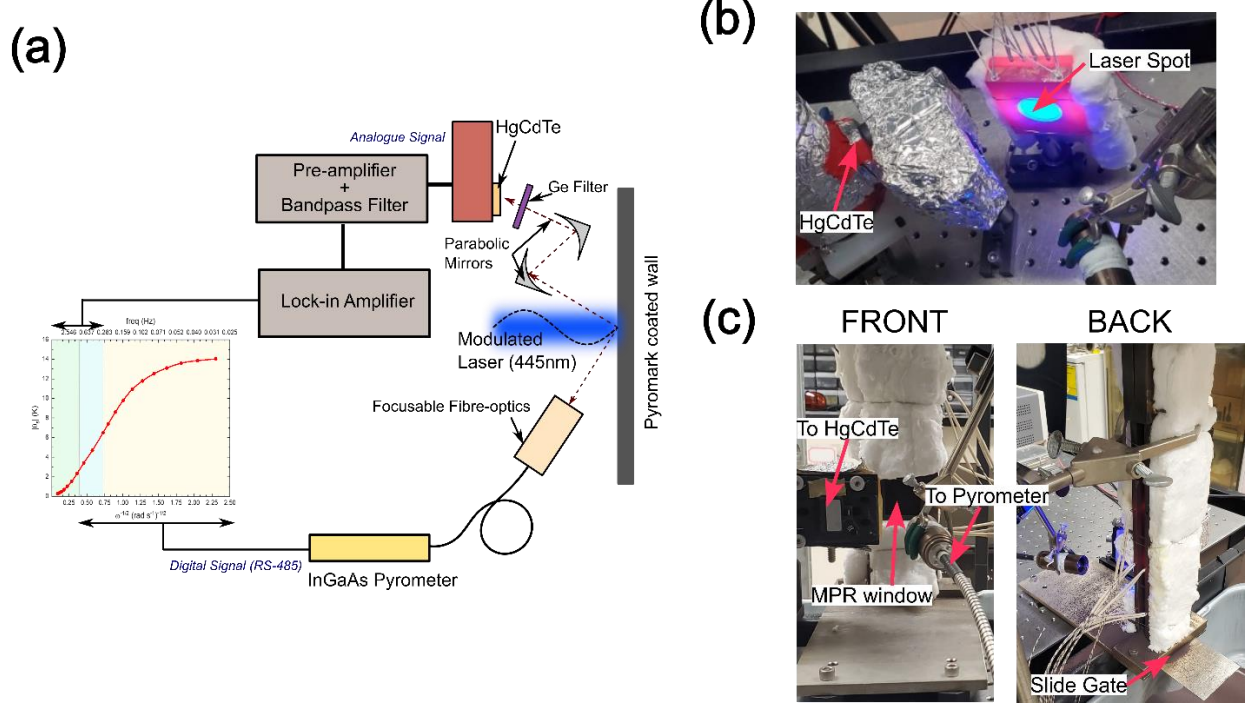

**Figure S1. (a)** Schematic of the MPR instrumentation; **(b)** Photograph of the holder for stationary packed particle MPR measurements; **(c)** Photographs of the front and back of the particle channel used for flowing particle MPR measurements.

Figure S1 (a) shows a schematic of the MPR instrumentation including the laser, infrared detectors, filter and lock-in amplifier. Figure S1 (b) and (c) are photographs of the stationary and flowing particles experiment setups, respectively. The front wall of the particles (both flowing and stationary) was a 100 μm thick metal sheet coated with a black light-absorbing paint (Pyromark 2500, LA-CO Industries, Inc.). A sinusoidal intensity modulated continuous-wave (455nm) laser was incident on the coated shim sheet creating the periodic heat source, and the surface temperature was recorded by a combination of infrared detectors. To maintain a 1-D heat transfer approximation, a ~22 mm diameter large laser spot was used and surface temperature was recorded from a 0.5 mm diameter spot around the centre. The laser heat flux was calibrated with the effusivity of a reference borosilicate glass sample.

The modulation frequency ranged between 20 to 0.03 Hz. The low frequency of 0.03 Hz enables a large thermal penetration depth and probes deeper into the particle bed. A maximum of 10 Hz was deemed sufficient as at higher frequencies, the penetration depth is limited to the sheet and coating, which is not of interest. In the high frequency range (>1 Hz), a MCT (HgCdTe) detector was used, and the voltage signal was fed to a lock-in amplifier to measure the amplitude. Due to the long settling times of the lock-in amplifier at low frequencies, a factory calibrated InGaAs pyrometer was used instead to record a digital temperature signal. This was then processed through a Fast Fourier Transform algorithm to extract the amplitude of temperature oscillations. The voltage from the MCT was referenced to this pyrometer temperature in an intermediate frequency range of 0.1-1 Hz and used to obtain the amplitude of surface temperature as a function of frequency over the entire frequency range. For complete details about the MPR methodology and instrumentation, please refer to [1,3,4].

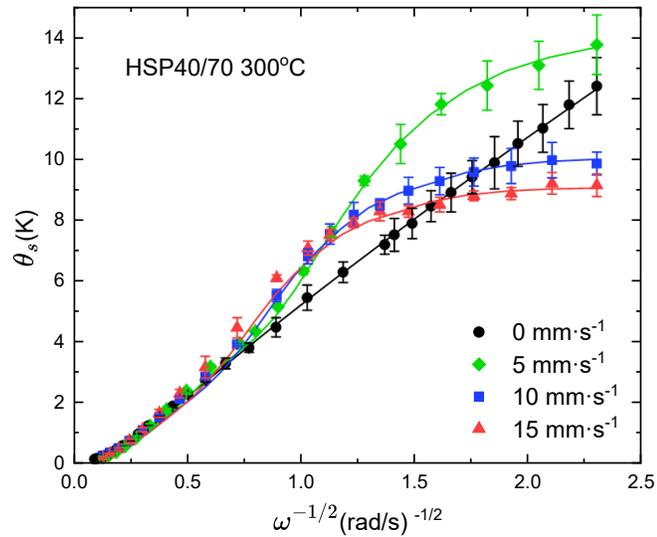

**Figure S2.** $|\theta_s|$ vs $\omega^{-1/2}$ data from MPR measurements on stationary and flowing HSP40/70 (404 µm) at 300 °C, as an example. Solids lines are model fitting results used to extract $k_{eff}$ and $D_{air}$. Error bar is standard deviation from three separate MPR measurements at the same conditions.

The $|\theta_s|$ vs $1/\sqrt{\omega}$ data of MPR measurements on HSP40/70 (404 µm) particles are shown as an example in Figure S2. For complicated geometries of our measurements, where the thermal penetration depth spans multiple sub-layers such as the near-wall and bulk regions of the particle beds (including the coating and wall in itself), an analytical solution can be used to fit the data and extract the near-wall thermal resistance and the effective bed thermal conductivity [2]. In line with earlier works, the near-wall thermal resistance was treated as an effective air-gap layer of $D_{air}$ thickness. The packed particle bed was treated as continuum medium with an effective bed thermal conductivity $k_{eff}$. The solution to the stationary particle bed $k_{eff}$ and $D_{air}$ can be obtained through a 1-D analytical model. A MATLAB implementation of the model in § 2.17 of Mandelis' textbook [2] was used to obtain $k_{eff}$ and $D_{air}$ via a non-linear least squares fit of stationary $|\theta_s|$ vs $1/\sqrt{\omega}$ data by the Trust-Region-Reflective algorithm.

For the flowing particle bed experiments, a 2-D (along depth of particle bed and direction of flow) COMSOL Multiphysics® model based on the 'Heat Transfer in Fluids and Solids' module was used. The governing equations in the fluid and solid domains are given by,

Fluid:
$$\rho c_p \frac{\partial T}{\partial t} + \rho c_p u_b \cdot \nabla T + \nabla \cdot (-k\nabla T) = q_0 \qquad (S2)$$

Solid:
$$\rho c_p \frac{\partial T}{\partial t} + \nabla \cdot (-k\nabla T) = q_0 \qquad (S3)$$

The flowing particle bed was treated as a homogeneous continuum with uniform thermophysical properties ($k_{\text{eff}}, \rho, c_p$) and a plug flow profile (no viscous boundary layer). A layer of air ($D_{\text{air}}$) was placed between the wall and particle bed to replicate the air-gap layer. Both $k_{\text{eff}}$ and $D_{\text{air}}$ are extracted from fitting the flowing $|\theta_s|$ vs $1/\sqrt{\omega}$ data using an iterative Levenberg-Marquardt algorithm.

A comparison between the COMSOL model and the analytical model is shown in Figure S3. First, a set of $k_{eff}$ and $D_{air}$ were input into the COMSOL model to generate an MPR curve; then the analytical model was used to fit the simulated MPR curve from the COMSOL model. The comparison was conducted at $300\,°C$, $500\,°C$, and $700\,°C$. The $k_{eff}$ and $D_{air}$ input into the COMSOL model are typical values for stationary CARBO ceramic particle beds measured via THW. It was found that both models can well predict the trend of MPR data for stationary beds. The analytical model has <4% lower fitted $k_{eff}$ compared to the COMSOL model while the fitted $D_{air}$ is <5% higher. This difference is well below the reduction in $k_{eff}$ (11-27%) and increase in $D_{air}$ (50-120%) observed in our measurements. Earlier works by the authors [5] using the analytical model fitting of MPR measurements of standard high temperature liquids such as paraffin wax, molten sulphur and molten salts, obtaining good agreement with literature thermal conductivity values.

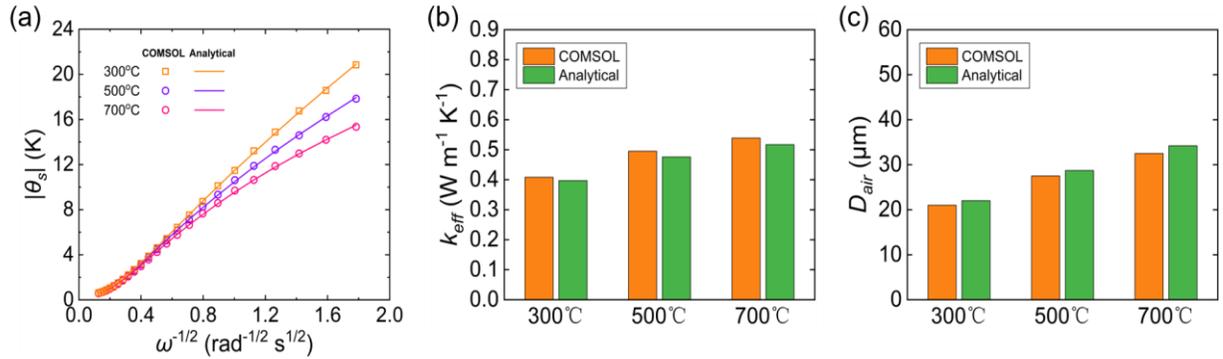

**Figure S3**. Comparison of COMSOL model and 1-D analytical model. (a) MPR curves, (b) $k_{eff}$, and (c) $D_{air}$ from both models.

Radiative surface heat losses from the coated MPR window were considered in both the 1D and 2D fitting models. While there is some heat loss due to natural convection, the contribution to total heat loss was around 20-5% over the measurement temperature range. Radiative HTC scales with T³ while convective HTC scales as T^(1/4). A verification of this assumption was conducted by repeating fitting with and without natural convection (radiation loss was present in both cases) for $10\,\text{mm} \cdot \text{s}^{-1}$ HSP 40/70 bed at $300°C$ and is shown in Figure S4 below.

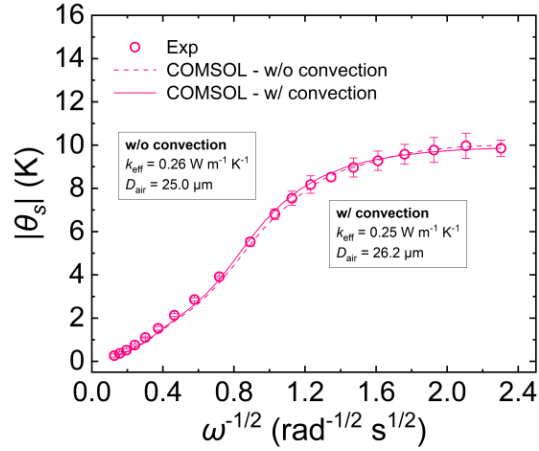

**Figure S4.** Comparison of COMSOL models without and with convective heat loss for moving HSP 40/70 bed at $10 \text{ mm} \cdot \text{s}^{-1}$ and $300 \text{ °C}$.

The addition of convective heat loss can lead to a 4% decrease in $k_{eff}$ and a 5% increase in $D_{air}$ at the lowest particle bed temperatures reported in this work. This error will be similar for both HSP40/70 and CP40/100 due to the similar values of $k_{eff}$, and lower for HSP16/30 due to higher $k_{eff}$. With increasing particle bed $k_{eff}$, since heat loss to ambient and heat flow into sample happen in parallel, the decrease in resistance to heat flow reduces the importance of heat loss. Similarly, at higher temperatures, the error in the fitting results introduced by convection should be similar or smaller than the values above. Neglecting the natural convection cause a small decrease in $k_{eff}$ and a small increase in $D_{air}$, thus the overall impact on the HTC calculation is minimal (a decrease of <4% for 10 mm·s-1 HSP 40/70 at 300°C). Complete information of the fitting models and procedure can be found in [1] and its accompanying Supplementary Information.

## S2. Error Analysis

### Uncertainty estimation of MPR measured $k_{\text{eff}}$ and $D_{\text{air}}$

The analytical $|\theta_s|$ is a nonlinear function of the laser modulation angular frequency $\omega$ and the interested parameter vector $\boldsymbol{P} = [k_{\text{eff}} \quad D_{\text{air}}]^T$

$$|\theta_s| = f(\omega, \boldsymbol{P}). \tag{S4}$$

In an MPR measurement, $n$ data points are sampled at angular frequency $\omega_i$ ($i = 1, 2, \ldots, n$). Here, the MPR-estimated $\boldsymbol{P}$ is denoted by $\widehat{\boldsymbol{P}} = [\hat{k}_{\text{eff}} \quad \widehat{D}_{\text{air}}]^T$. The uncertainty in $\widehat{\boldsymbol{P}}$, namely $\Delta \boldsymbol{P}$, can be calculated using a Taylor expansion of the function $f$ around $\widehat{\boldsymbol{P}}$ [6]:

$$\Delta \boldsymbol{P} = (\bar{\boldsymbol{J}}^T \bar{\boldsymbol{J}})^{-1} \bar{\boldsymbol{J}}^T \Delta \boldsymbol{\Theta}, \tag{S5}$$

where $\bar{\boldsymbol{J}}$ is the Jacobian matrix

$$\bar{\boldsymbol{J}} = \begin{bmatrix} \left.\dfrac{\partial f(\omega_1, \boldsymbol{P})}{\partial k_{\text{eff}}}\right|_{\widehat{\boldsymbol{P}}} & \left.\dfrac{\partial f(\omega_1, \boldsymbol{P})}{\partial D_{\text{air}}}\right|_{\widehat{\boldsymbol{P}}} \\ \vdots & \vdots \\ \left.\dfrac{\partial f(\omega_n, \boldsymbol{P})}{\partial k_{\text{eff}}}\right|_{\widehat{\boldsymbol{P}}} & \left.\dfrac{\partial f(\omega_n, \boldsymbol{P})}{\partial D_{\text{air}}}\right|_{\widehat{\boldsymbol{P}}} \end{bmatrix}, \tag{S6}$$

and $\Delta \boldsymbol{\Theta}$ is the uncertainty vector of $|\theta_s|$ with two major sources: the uncertainty from the measurement and the error in fitting, which is expressed as

$$\Delta \boldsymbol{\Theta}_i = \sqrt{|\theta_s|_{\text{std},i}^2 + \left(|\theta_s|_{\text{exp},i} - |\theta_s|_{\text{fit},i}\right)^2}, \tag{S7}$$

where $|\theta_s|_{\text{exp},i}$, $|\theta_s|_{\text{std},i}$, and $|\theta_s|_{\text{fit},i}$ are the raw data, the measurement standard deviation (error bar in Figure S2), and the COMSOL-simulated value at $\omega_i$, respectively. The uncertainties of the MPR-estimated $k_{\text{eff}}$ and $D_{\text{air}}$ are given by the diagonal elements of the covariance matrix $E[\Delta \boldsymbol{P} \Delta \boldsymbol{P}^T]$. The elements in $\bar{\boldsymbol{J}}$ are approximated by

$$\left.\frac{\partial f(\omega_i, \boldsymbol{P})}{\partial k_{\text{eff}}}\right|_{\widehat{\boldsymbol{P}}} = \frac{|\theta_s|_{\text{fit},i}\big|_{[\hat{k}_{\text{eff}} + \Delta k_{\text{eff}} \quad \widehat{D}_{\text{air}}]^T} - |\theta_s|_{\text{fit},i}\big|_{[\hat{k}_{\text{eff}} \quad \widehat{D}_{\text{air}}]^T}}{\Delta k_{\text{eff}}}, \tag{S8}$$

and

$$\left.\frac{\partial f(\omega_i, \boldsymbol{P})}{\partial D_{\text{air}}}\right|_{\widehat{\boldsymbol{P}}} = \frac{|\theta_s|_{\text{fit},i}\big|_{[\hat{k}_{\text{eff}} \quad \widehat{D}_{\text{air}} + \Delta D_{\text{air}}]^T} - |\theta_s|_{\text{fit},i}\big|_{[\hat{k}_{\text{eff}} \quad \widehat{D}_{\text{air}}]^T}}{\Delta D_{\text{air}}}, \tag{S9}$$

where $\Delta k_{\text{eff}}$ and $\Delta D_{\text{air}}$ are 1% of $\hat{k}_{\text{eff}}$ and $\widehat{D}_{\text{air}}$, respectively.

**Error propagation of $\delta k_{\text{eff}}$ and $\delta D_{\text{air}}$**

Once the error in $k_{\text{eff}}$ and $D_{\text{air}}$ is obtained, their respective errors $\delta k_{\text{eff}}$ and $\delta D_{\text{air}}$ are then propagated as

$$\delta Pe_{D_h} = Pe_{D_h} \frac{\delta k_{\text{eff}}}{k_{\text{eff}}} \tag{S10}$$

$$\delta \left(\frac{R_{\text{nw}}}{R_p}\right) = \left(\frac{R_{\text{nw}}}{R_p}\right)\left(\frac{\delta D_{\text{air}}}{D_{\text{air}}} + \frac{\delta k_{\text{eff}}}{k_{\text{eff}}}\right) \tag{S11}$$

Error is the averaged Nusselt number is approximated as the error in the fully developed Nusselt number as

$$\delta \overline{Nu_{D_h}} = \left(\overline{Nu_{D_h}}\right)^2 \left(\frac{k_{\text{eff}}}{k_{\text{air}} D_h} \delta D_{\text{air}} + \frac{D_{\text{air}}}{k_{\text{air}} D_h} \delta k_{\text{eff}}\right) \tag{S12}$$

Where

$$\overline{Nu_{D_h}} = \left(\frac{1}{12} + \frac{R_{\text{nw}}}{4R_p}\right)^{-1} \tag{S13}$$

Similarly, the error in the average particle-to-wall heat transfer coefficient is given as

$$\delta \bar{h} = (\bar{h})^2 \left(\frac{1}{4k_{\text{air}}} \delta D_{\text{air}} + \frac{D_h}{12 k_{\text{eff}}^2} \delta k_{\text{eff}}\right) \tag{S14}$$

**References:**

[1] Zhang, X., Adapa, S., Feng, T., Zeng, J., Chung, K. M., Ho, C., Albrecht, K., and Chen, R., 2024, "Micromechanical Origin of Heat Transfer to Granular Flow," Phys. Rev. E, **109**(4), p. L042902.
[2] Mandelis, A., 2001, *Diffusion-Wave Fields*, Springer New York, New York, NY.
[3] Zeng, J., Chung, K. M., Wang, Q., Wang, X., Pei, Y., Li, P., and Chen, R., 2021, "Measurement of High-Temperature Thermophysical Properties of Bulk and Coatings Using Modulated Photothermal Radiometry," International Journal of Heat and Mass Transfer, **170**, p. 120989.
[4] Zeng, J., Chung, K. M., Adapa, S. R., Feng, T., and Chen, R., 2021, "In-Situ Thermal Transport Measurement of Flowing Fluid Using Modulated Photothermal Radiometry," International Journal of Heat and Mass Transfer, **180**, p. 121767.
[5] Chung, K. M., Feng, T., Zeng, J., Adapa, S. R., Zhang, X., Zhao, A. Z., Zhang, Y., Li, P., Zhao, Y., Garay, J. E., and Chen, R., 2023, "Thermal Conductivity Measurement Using Modulated Photothermal Radiometry for Nitrate and Chloride Molten Salts," International Journal of Heat and Mass Transfer, **217**, p. 124652.
[6] Yang, J., Ziade, E., and Schmidt, A. J., 2016, "Uncertainty Analysis of Thermoreflectance Measurements," Review of Scientific Instruments, **87**(1), p. 014901.